\documentclass[prl,aps,twocolumn,amsmath,amssymb,floatfix,reprint,footinbib,superscriptaddress]{revtex4-2}
\usepackage{graphics}      
\usepackage{graphicx}      
\usepackage{longtable}     
\usepackage{url}           
\usepackage{bm}    
\usepackage{braket}        
\usepackage{xcolor}
\usepackage{comment}
\usepackage{siunitx}
\usepackage[T1]{fontenc}
\usepackage[title]{appendix}
\usepackage{amsmath,amssymb,epsfig,color}
\usepackage{mathrsfs}
\usepackage{amsfonts}
\usepackage{dcolumn}
\usepackage{subfigure}
\usepackage{soul}
\usepackage{array}
\usepackage{siunitx}
\usepackage{tabularx}
\usepackage[colorlinks,linkcolor=blue,citecolor=blue,hyperindex,bookmarks=false,pdfstartview=FitH]{hyperref}
\usepackage[left]{lineno}  

\DeclareMathOperator*{\argmin}{argmin}

\newcommand{\beq}{\begin{equation}}
\newcommand{\eeq}{\end{equation}}
\newcommand{\beqa}{\begin{eqnarray}}
\newcommand{\eeqa}{\end{eqnarray}}

\newcommand{\llangle}{\langle\!\langle}
\newcommand{\rrangle}{\rangle\!\rangle}

\newcommand{\figref}[1]{\mbox{Fig.~\ref{#1}}}

\newcommand{\figpanel}[2]{Fig.~\hyperref[#1]{\ref*{#1}(#2)}}
\newcommand{\figpanels}[3]{Fig.~\hyperref[#1]{\ref*{#1}(#2)-(#3)}}
\newcommand{\figpanelNoPrefix}[2]{\hyperref[#1]{\ref*{#1}(#2)}}

\usepackage{mleftright} 
 
\begin{document}
\title{Quantum Process Tomography with Digital Twins of Error Matrices}

\author{Tangyou Huang}
\affiliation{Department of Microtechnology and Nanoscience, Chalmers University of Technology, 41296 Gothenburg, Sweden}
\affiliation{QTF Centre of Excellence and InstituteQ, Department of Applied Physics, Aalto University, 00076 Aalto, Finland}

\author{Akshay Gaikwad}
\email{akshayga@chalmers.se}
\affiliation{Department of Microtechnology and Nanoscience, Chalmers University of Technology, 41296 Gothenburg, Sweden}

\author{Ilya Moskalenko}
\affiliation{QTF Centre of Excellence and InstituteQ, Department of Applied Physics, Aalto University, 00076 Aalto, Finland}

\author{Anuj Aggarwal}
\author{Tahereh Abad}
\affiliation{Department of Microtechnology and Nanoscience, Chalmers University of Technology, 41296 Gothenburg, Sweden}
\author{Marko Kuzmanovi\'c}
\author{Yu-Han Chang}
\author{Ognjen Stanisavljevi\'c}
\affiliation{QTF Centre of Excellence and InstituteQ, Department of Applied Physics, Aalto University, 00076 Aalto, Finland}

\author{Emil Hogedal}
\author{Christhopher Warren}
\author{Irshad Ahmad}
\author{Janka Biznárová}
\author{Amr Osman}
\author{Mamta Dahiya}
\author{Marcus Rommel}
\author{Anita Fadavi Rousari}
\author{Andreas Nylander}
\author{Liangyu Chen}
\affiliation{Department of Microtechnology and Nanoscience, Chalmers University of Technology, 41296 Gothenburg, Sweden}

\author{Jonas Bylander}
\affiliation{Department of Microtechnology and Nanoscience, Chalmers University of Technology, 41296 Gothenburg, Sweden}

\author{Gheorghe Sorin Paraoanu}
\affiliation{QTF Centre of Excellence and InstituteQ, Department of Applied Physics, Aalto University, 00076 Aalto, Finland}

\author{Anton Frisk Kockum}
\author{Giovanna Tancredi}
\affiliation{Department of Microtechnology and Nanoscience, Chalmers University of Technology, 41296 Gothenburg, Sweden}

\date{\today }
	
\begin{abstract}
Accurate and robust quantum process tomography (QPT) is crucial for verifying quantum gates and diagnosing implementation faults in experiments aimed at building universal quantum computers. However, the reliability of QPT protocols is often compromised by faulty probes, particularly state preparation and measurement (SPAM) errors, which introduce fundamental inconsistencies in traditional QPT algorithms. 
We propose and investigate enhanced QPT for multi-qubit systems by integrating the error matrix in a digital twin of the identity process matrix, enabling statistical refinement of SPAM error learning and improving QPT precision.
Through numerical simulations, we demonstrate that our approach enables highly accurate and faithful process characterization. We further validate our method experimentally in superconducting quantum processors, achieving at least an order-of-magnitude fidelity improvement over standard QPT.
Our results provide a practical and precise method for assessing quantum gate fidelity and enhancing QPT on a given hardware.
\end{abstract}

\maketitle

\twocolumngrid
\textit{Introduction.}
Significant advancements have been made in building large-scale quantum processors using diverse physical platforms~\cite{timo-nat-2025, bluvstein-nat-2024, moses-prx-2023, arute-nat-2019}. Although a higher qubit count provides exponential computational benefits, it also brings major challenges in implementing high-fidelity multi-qubit gates and accurately characterizing them for further enhancement~\cite{hendrik-prapp-2020, tahereh-prl-2022, tahereh-quantum-2025, chris-npj-2023, robin-quantum-2021, roth-prxquantum-2023, riofrio-2017}. Identifying errors in gate implementation and improving quantum architectures require more than a single scalar measure, such as gate fidelity from randomized benchmarking protocols~\cite{robin-prl-2017, robin-nrp-2025, robin-arxiv-2025, robin-arxiv-2024}. Instead, a comprehensive characterization of the entire quantum process is essential, which is typically achieved through quantum process tomography (QPT)~\cite{chuang-jmo-09, white-pra-2004, zoller-prl-1997}.

Quantum process tomography involves preparing a set of known input states $\{ \rho_i \}$, applying a completely positive and trace-preserving (CPTP) quantum process $\mathcal{E}$, and measuring a set of observables $\{ M_{\mu} \}$, typically chosen as elements of a positive operator-valued measure (POVM). This yields a collection of measurement outcomes: $p_{i,\mu} = \text{Tr}\left[M_{\mu}\mathcal{E}(\rho_i)\right]$.
In experiments, this procedure demands high-precision state preparation and measurement (SPAM) in order to faithfully characterize the underlying quantum process---a requirement that remains challenging on state-of-the-art hardware platforms~\cite{robin-nrp-2025}.
Nevertheless, standard QPT assumes ideal probes $\rho_i$ and $M_\mu$ in SPAM operations, and applies a post-processing algorithm $\mathcal{J}$ to noisy data points $\tilde{p}_{i,\mu}$:
\beq\label{eq:std-QPT}
\text{std-QPT}: \quad \mathcal{J}( \rho_i,  M_{\mu},  \tilde{p}_{i,\mu}) \to \tilde{\chi}.
\eeq
This leads to internal inconsistencies~\cite{self-inconsistency}: the noisy measurement outcomes $\{ \tilde{p}_{i,\mu} \}$ are incorrectly attributed to ideal SPAM operations, thereby distorting the reconstruction and interpretation of the resulting process matrix $\tilde{\chi}$. As a result, standard QPT frequently yields unreliable or even misleading characterizations. 
This fundamental issue of SPAM-induced self-inconsistency in QPT was first recognized and systematically analyzed over a decade ago~\cite{merkel-pra-2013, stark-pra-2014, robin-arxiv-2013, james-conference-2013}, prompting the development of gate set tomography (GST)~\cite{greenbaum-arxiv-2015, robin-quantum-2021, li-qst-2024, cao-prl-2024, erik-qst-2020, vinas-nature-2025}. GST is a protocol that enables SPAM-error-free characterization of quantum gate sets. However, its experimental and computational overhead is substantially higher than that of QPT, rendering it impractical for systems beyond two qubits~\cite{roth-prxquantum-2023}. 

\begin{figure} 
\centering
\includegraphics[width=\columnwidth]{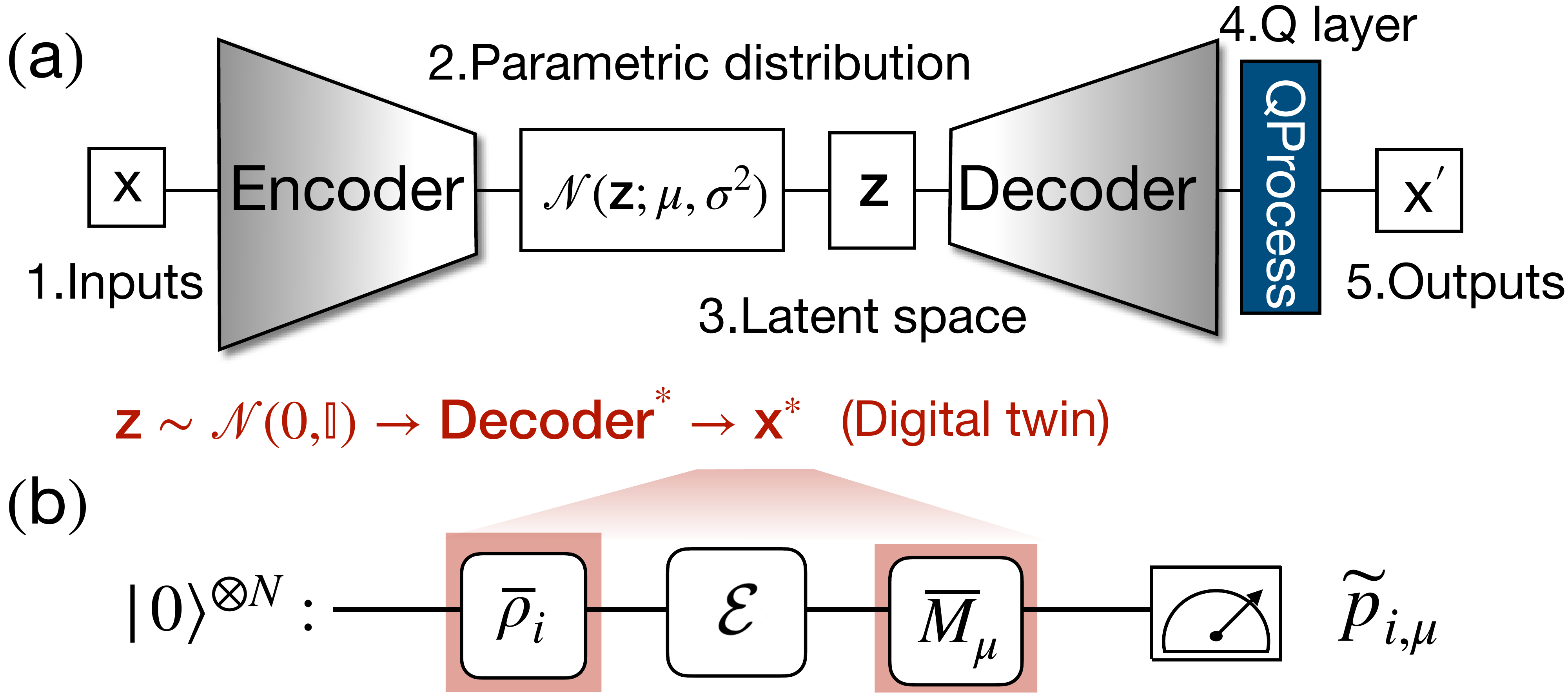}
\caption{\textbf{Digital twin-enhanced quantum process tomography.} 
    \textbf{(a)} The variational autoencoder (VAE) consists of an encoder and a decoder built with deep neural networks. The input training data \(\textbf{x}\) is mapped by the encoder into a parametric probability distribution \(\mathcal{N}(\textbf{z}; \mu, \sigma^2)\). The latent variable \(\textbf{z}\) is sampled from this distribution and used to reconstruct the output \(\textbf{x}'\) through the decoder and a pre-designed quantum processing layer (QProcess).  
    \textbf{(b)} The digital twin is applied to reconstruct the error matrix $\textbf{x}^* \to \chi^{I}_{\text{DT}}$ using a trained VAE, 
    enhancing EM-QPT for a quantum process $\mathcal{E}$.}
\label{fig:figure1}
\end{figure}

In this Letter, we propose a generic framework to realize a nearly self-consistent QPT protocol for multi-qubit systems (see \figref{fig:figure1}).
By reconstructing effective probes $\{ \bar{\rho}_i, \bar{M}_{\mu} \}$, our SPAM-error-mitigated QPT (EM-QPT) achieves significantly higher accuracy compared to the standard, self-inconsistent QPT technique. To further improve precision and robustness, we incorporate a machine learning (ML) approach~\cite{VAE0,huang2022prA,huang2024quantum} that learns the statistical feature of SPAM errors by constructing a digital twin of identity process matrices. Remarkably, we validate our method experimentally in two superconducting quantum processors, achieving at
least an order-of-magnitude fidelity improvement over standard QPT. Our results establish a scalable and practical framework for high-precision quantum diagnostics, broadly applicable to quantum computing, benchmarking, and control.

\textit{Error-mitigated QPT.} 
Our objective is to precisely estimate the process matrix of an arbitrary quantum operation $\mathcal{E}$ while accounting for SPAM errors~\cite{blumekohout2024easybetterquantumprocess, Korotkov09pra,korotkov2013error,duan2020PRA}. 
We utilize identity QPT: only performing state preparation $\mathcal{E}_{\text{sp}}$ and measurement $\mathcal{E}_{\text{m}}$, yielding $\chi^{\text{I}} \equiv \mathcal{E}_{\text{m}}\circ \mathcal{E}_{\text{sp}}$. Ideally, the identity process matrix will be $\chi^{\text{I}}_{mn} = \delta_{m0}\delta_{n0}$, where $\delta_{mn}$ is the Kronecker delta. Deviations from the ideal $\chi^{\text{I}}$ indicate the presence of SPAM errors in the experiment, resulting in $\tilde{\chi}^I$, which we refer to as an \textit{`error matrix'}~\cite{korotkov2013error}. By changing the argument in the QPT algorithm, we can determine the noisy input states and measurement operators~\cite{SM}:
\begin{equation}
\mathcal{J}[\{\rho_i\}, \{M_{\mu}\}, \tilde{\chi}^I] \rightarrow \{\bar{\rho}_{i}\}, \{\bar{M}_{\mu}\}. \label{eq:rho_tilde}
\end{equation}
When computing $\{ \bar{\rho}_i \}$, we assume ideal measurement operators, and vice versa (standard quantum state and detector tomography with the error matrix), since gauge symmetry due to unitary invariance~\cite{SM} prevents simultaneously determining $\{ \bar{\rho}_i, \bar{M}_{\mu} \}$ with arbitrary accuracy and precision~\cite{laflamme-prr-2021, robin-natcom-2017, karol-quantum-2018, lin-njp-2019}. We note that a recent theoretical work~\cite{blumekohout2024easybetterquantumprocess} proposes a strategy similar to EM-QPT, in which the probes are revised by leveraging prior knowledge of the error matrix.

Here, we benchmark the practical, error-mitigated, and nearly self-consistent version of QPT: 
\beq\label{eq:EM-QPT}
\text{EM-QPT}: \mathcal{J}(\bar{\rho}_i, \bar{M}_{\mu}, \tilde{p}_{i,\mu} ; \tilde{\chi}^I).
\eeq

This EM-QPT approach is resource-intensive, particularly in applications where frequent process characterization is required, such as gate optimization~\cite{QOC2014RB,RL2021prxq}.
It is also potentially fragile in the presence of anomaly errors, such as glitches in experiments.
Since the error matrix $\tilde{\chi}^I$ is independent of the process to be characterized, it is natural to explore whether ML techniques can be leveraged to learn the statistical behavior of SPAM errors hidden in $\tilde{\chi}^I$, for more efficient error mitigation.

Inspired by a recent study~\cite{huang2024quantum}, we use a generative model as a digital twin of the error matrix to enhance EM-QPT. We find that the digital twin, a trained deep neural network, effectively captures the underlying characteristics of SPAM errors, yielding a more refined version of Eq.~\eqref{eq:rho_tilde}. It can potentially outperform real-time error-matrix acquisition, enabling high-precision QPT with more robust and efficient SPAM-error mitigation.

\begin{figure} 
\centering
\includegraphics[width=\columnwidth]{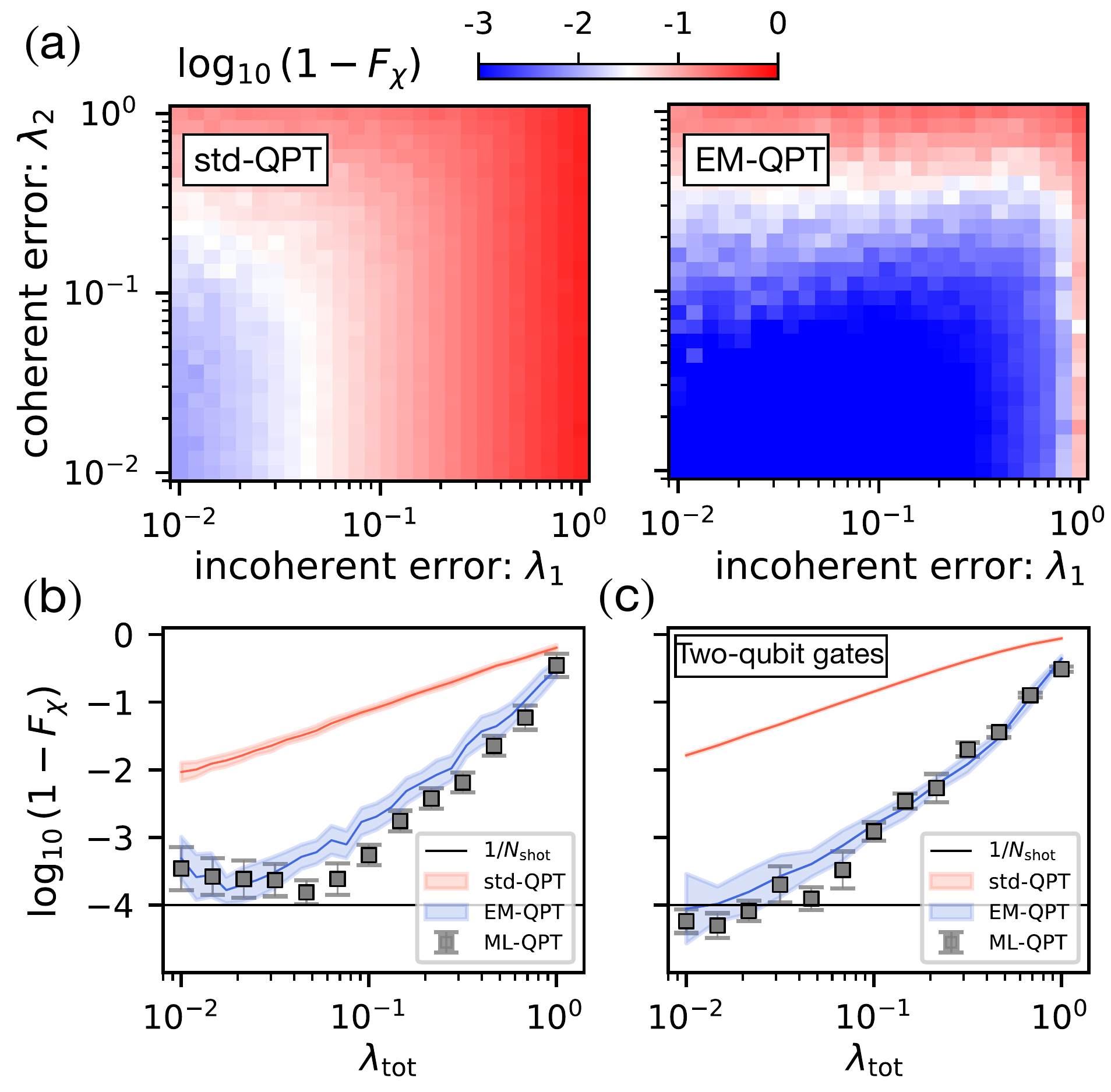}
\caption{\textbf{Numerical results for EM-QPT of one- and two-qubit gates.} (a) Average process infidelity of single-qubit gates using std-QPT {(left panel)} and EM-QPT {(right panel)}, as a function of coherent and incoherent errors.
(b, c) Fidelity for the evenly mixed error regime 
for single- and two-qubit gates, respectively. Solid curves and squares are the average results over $10^2$ gate samples, with shadow and error bar showing one standard deviation.
The horizontal line is the statistical error $1/N_{\text{shot}}$ with the shot number $N_{\text{shot}}=10^4$.}
\label{fig:numeric results}
\end{figure}

\textit{Digital twin of the error matrix.}
Our generative model to construct the digital twin of the error matrix is a variational autoencoder (VAE)~\cite{VAE0}, which integrates deep learning with probabilistic frameworks to learn a latent representation of training data (see \figref{fig:figure1}).
The VAE consists of an \textit{encoder} that maps the input \( \textbf{x} \) to a latent vector $\textbf{z}$ obeying a probability distribution \( \mathbb{Q}(\textbf{z}|\textbf{x}) \). The latent vector is a numerical representation of the essential features of the input data, usually in a lower-dimensional space.
The \textit{decoder} reconstructs the input data \( \textbf{x}\to\textbf{x}' \) from a sampled latent vector $\textbf{z}$. 
Both \textit{encoder} and \textit{decoder} are two deep neural networks~\cite{SM}. 
To ensure that the VAE output is CPTP, we introduce a {\tt QProcess} layer~\cite{SM} using Cholesky decomposition~\cite{PhysRevA.61.010304,Shahnawaz2021cGAN}. See the workflow in \figref{fig:figure1}.

We characterize the SPAM errors by constructing the digital twin of the error matrix $\tilde{\chi}^I$.
In practice, we first collect a training database $\textbf{X} = \{\textbf{x}^{(i)}\}_{i=0}^{N_x}$ of $N_{x}$ independent QPT experiments for the identity process, which is implemented by applying a short idle time of a few nanoseconds in experiments~\cite{SM}. 
The digital twin of the error matrix, $\textbf{x}' \to \chi^I_{\rm DT}$, is expected to statistically mimic the major pattern of SPAM errors embedded in the error matrix. In this vein, we utilize the digital twin to perform the EM-QPT protocol; the digital twin is applicable to an arbitrary quantum process, as the error matrix is independent of the gate operation under test. We thus introduce the machine learning-enhanced QPT, 
\begin{align}
\text{ML-QPT}&: \mathcal{J}(\bar{\rho}_i, \bar{M}_{\mu}, \tilde{p}_{i,\mu}; \chi^{I}_{\text{DT}}),  
\end{align}
where $\chi^{I}_{\text{DT}}$ is the digital twin of the error matrix.
See Appendix D in~\cite{SM} for detailed information about the model structure and learning process.

\textit{Numerical simulation}.
In standard $N$-qubit QPT, each qubit is first initialized in an initial state
by applying a gate $U_1 \in \{\texttt{I}, \texttt{R}_x(-\pi/2), \texttt{R}_y(-\pi/2), \texttt{X}\}$. Then, the quantum gate $\mathcal{G}$ under investigation is applied. To enable full process characterization, a set of informationally complete rotation gates $U_2 \in \{\texttt{I}, \texttt{R}_x(\pi/2), \texttt{R}_y(\pi/2)\}$ are used prior to measuring the qubits in the computational basis.
In practice, the noisy readout is composed as
\beq
\tilde{p}_{i,\mu}=
\llangle M_\mu|\mathcal{E}_{\rm m}^{\dagger}\circ \mathcal{E}_{\rm gate}(\mathcal{G})\circ\mathcal{E}_{\rm sp}|\rho_i\rrangle,
\eeq
where $|\hat{O}\rrangle$ denotes the column-vector form of the operator $\hat{O}$, and $\mathcal{E}_{\rm m}$, $\mathcal{E}_{\rm gate}$, $\mathcal{E}_{\rm sp}$ represent the error channels acting on the measurement, gate, and initial states, respectively.
To simulate incoherent SPAM errors, we use depolarizing error channel $\mathcal{E}_{\rm dep}(\rho;\lambda ) = (1-\lambda)\rho+\frac{\lambda}{2^N}I$, where $I$ is the identity operator.
We randomly sample the error strength for state preparation $\mathcal{E}_{\rm sp} = \mathcal{E}_{\rm dep}(\rho;\lambda_{\rm sp} )$ and measurements $\mathcal{E}_{\rm m} =\mathcal{E}_{\rm dep}(\rho;\lambda_{\rm m})$ in terms of a given error rate $\lambda_{ 1}=\lambda_{\text{m}}+\lambda_{\text{sp}}$. 
We also introduce coherent errors with a unitary channel $\mathcal{E}_{\rm uni} (\rho)= U\rho U^\dagger$ by adding a rotation shift $\Delta\theta = \theta'-\theta_0$ on a rotation gate ${\tt R_{i\in[x,y,z]}(\theta_0)}\to {\tt R_{i\in[x,y,z]}(\theta')}$ in SPAM; we uniformly sample the deviation $ \Delta\theta/\pi \in  [-\lambda_2,\lambda_2]$ ($\lambda_2\in[0,1]$). Therefore, we express the total SPAM error as
$\lambda_{\text{tot}} = (\lambda_1+\lambda_2)/2 \in[0,1]$.
The numerical experiment consists of three steps:
(i) set $\mathcal{G} = {\tt I}$, the $N$-qubit identity gate, and perform std-QPT to obtain the error matrix $\tilde{\chi}^I$;  
(ii) reconstruct $4^N$ noisy quantum states and $6^N$ observables using Eq.~\eqref{eq:rho_tilde};
(iii) perform std-QPT on a randomly selected unitary gate $\mathcal{G}$ using the error-mitigated probes, giving a corrected process matrix according to Eq.~\eqref{eq:EM-QPT}. Here, we focus on unitary operations, but the EM-QPT approach is valid for any general CPTP process~\cite{SM}.

In \figref{fig:numeric results}, we present numerical results for std-QPT, EM-QPT, and ML-QPT under the influence of both incoherent and coherent errors. Each data point represents the average process infidelity~\cite{fidelity-definition} computed over $10^2$ randomly chosen unitary gates. Particularly, in \figpanel{fig:numeric results}{a}, we analyze the gate infidelity as a function of coherent error $\lambda_2$ and incoherent error $\lambda_1$ for a single-qubit system. EM-QPT outperforms std-QPT with significant fidelity improvement. Furthermore, we investigate how the infidelity scales with the total error $\lambda_{\rm tot}$ in the case of evenly mixed contributions, i.e., $\lambda_1 = \lambda_2$, with the corresponding results for one- and two-qubit gates presented in \figpanel{fig:numeric results}{b} and \figpanelNoPrefix{fig:numeric results}{c}, respectively.
For ML-QPT, we collected $10^3$ error matrices of each data point to train a digital twin model across a range of $\lambda_{\text{tot}}$, and achieved better performance than with EM-QPT, as seen in \figpanels{fig:numeric results}{b}{c}.
Moreover, we also demonstrate that our method maintains high performance even in the presence of extremely biased SPAM errors  $(\frac{\lambda_{\rm m}}{\lambda_{\rm sp}}\to \infty /0)$; see Appendix H in~\cite{SM}.

\begin{figure} 
\centering
\includegraphics[width=1\columnwidth]{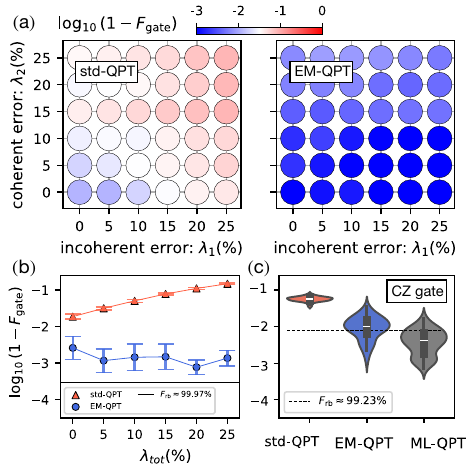} 
\caption{\textbf{Experimental results for one- and two-qubit gates.} 
(a) Performance of std-QPT (left) and EM-QPT (right) as a function of coherent and incoherent errors.
(b) The points from (a) with $\lambda_1=\lambda_2$.
Each data point in (a)/(b) is averaged over $15$/$10^2$ QPT experiments for all 24 single-qubit Clifford gates.
(c) Infidelity distribution over $10^2$ QPT experiments of a CZ gate estimated with std-QPT, EM-QPT, and ML-QPT. The inner box plot indicates the median (white horizontal line) and the interquartile range (black box). Here, SPAM errors $\sim \SI{6}{\percent}$ and the RB fidelity is $\SI{99.23}{\percent}$.
}
\label{fig:gate_exp}
\end{figure}

Next, we verify EM-QPT in experiments with single-qubit Clifford gates.  
We consider the average gate fidelity~\cite{PhysRevA.60.1888,nielsen2002simple,PhysRevLett.102.090502}
\beq\label{eq:gate fidelity}
\mathcal{F}_{\text{gate}}=\frac{d\mathcal{F}_\chi+1}{d+1},
\eeq
where $d = 2^N$ is the Hilbert-space dimension of the $N$-qubit system; the process fidelity $\mathcal{F}_\chi$~\cite{fidelity-definition} is obtained through QPT.
For small gate errors, randomized benchmarking (RB)~\cite{RB2011prl} statistically captures the average gate error over Clifford gates, implying that the RB fidelity then approximates the average gate fidelity: $\mathcal{F}_{\text{rb}} \approx \mathcal{F}_{\text{gate}}$.  In the following, we demonstrate the experimental implementations of our method.

\textit{Experiments on single-qubit gates.} We implement our method on 24 single-qubit Clifford gates on a superconducting quantum processor (see Device A in \cite{SM}). 
In experiments, we calibrate both single-qubit gates, achieving fidelity of $99.96\%$ using RB measurements, and readout performance~\cite{SM}. 
We then introduce coherent and incoherent errors. Incoherent errors are introduced by biasing the optimized amplitude of square pulses $A_{r0}$, used for readout, by $\lambda_1$: $A_r =(1- \lambda_1)A_{r0}$ ($\lambda_1\in\mathbb{R}$). Coherent errors are introduced by adding rotation uncertainties to the single-qubit gates through amplitude fluctuations: 
the target amplitude $A_0$ is modified to $A_U = (1+r) A_0$, with the offset $r$ uniformly sampled as $r \sim \mathcal{U}(-\lambda_2, \lambda_2)$ ($\lambda_2 \in [0, 1]$). See the End Matter for details about the error setup.

In \figpanels{fig:gate_exp}{a}{b}, we present experimental results for std-QPT and EM-QPT for single-qubit Clifford gates under varying levels of coherent and incoherent SPAM errors; EM-QPT yields at least an order-of-magnitude improvement in gate fidelity. 
Notably, RB outperforms EM-QPT because SPAM errors cannot be explicitly separated from the error matrix. We leave further optimization of the QPT method~\cite{blumekohout2024easybetterquantumprocess} through adjusting the weight function between state preparation and measurements for future work.

\textit{Two-qubit CZ gates.} We start with a well-tuned adiabatic CZ gate with \( \mathcal{F}^{\rm CZ}_\text{rb} \sim \SI{99.23}{\percent} \) (see Device B in \cite{SM}). 
After fine-tuned calibration with single-qubit gate fidelity $\mathcal{F}_{\rm rb}\approx 99.95\%$~\cite{SM}, we observed a $6\%$ process fidelity reduction due to SPAM errors.
We perform identity QPT followed by QPT of this gate $10^2$ times. We can thus obtain the process fidelity of the CZ gate with both EM-QPT and ML-QPT.
In \figpanel{fig:gate_exp}{c}, we present the probability distribution of gate infidelity estimated by Eq.~\eqref{eq:gate fidelity} using std-QPT, EM-QPT, and ML-QPT; the ML-QPT results are based on a digital twin trained on the $10^2$ error matrices. As a result, the gate fidelities estimated by EM-QPT and ML-QPT are significantly closer to the RB fidelity than those from standard QPT, with the overestimation in ML-QPT attributed to the limited size of the training dataset.

\textit{Precision and sensitivity.}
In experiments with our method, the gate fidelity of an unknown gate $\mathcal{G}$ can be statistically estimated over the set $\mathcal{S}_I=\{\tilde{\chi}^{I}_{1},\tilde{\chi}^{I}_{2},...,\tilde{\chi}^{I}_{N_{\text{err}}}\}$, forming a probability distribution $\mathbb{P}_{\text{EM}}(\mathcal{F}(G;\tilde{\chi}^{I}))|_{\tilde{\chi}^{I}\in\mathcal{S}_I}$, where the variance is primarily induced by SPAM errors.
However, averaging fidelity over all error matrices may reduce EM-QPT precision due to experimental anomalies. To address this, we use ML to extract the dominant SPAM error patterns, creating a digital twin that reconstructs them.
Therefore, the gate fidelity relying on the digital twin admits the distribution $\mathbb{P}_{\text{ML}}(\mathcal{F})|_{{\tt D^*(\textbf{z})}\to\chi^I }$, where ${\tt D^*}(\textbf{z})|_{\textbf{z}\sim\mathcal{N}(0,\mathbb{I})}$ is the decoder from the trained VAE model.
\begin{figure}
\centering
\includegraphics[width=1\columnwidth]{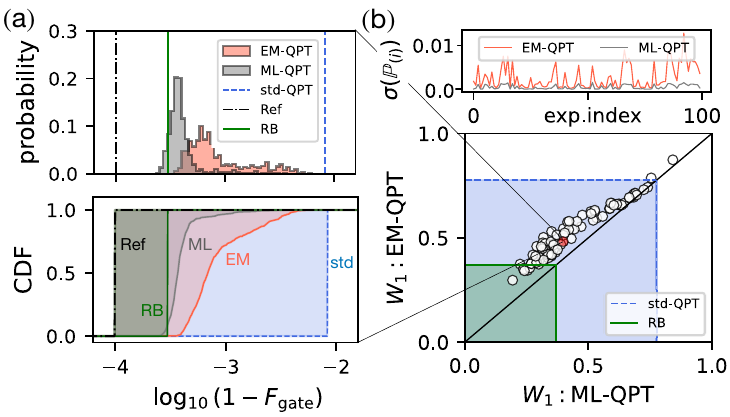}
\caption{\textbf{Precision and sensitivity.} 
(a) Top: infidelity distribution of EM-QPT and ML-QPT, based on the training dataset and digital twin, compared with std-QPT (dashed blue), reference (dot-dashed black), and RB (solid green). 
Bottom: CDFs used to calculate the $W_1$ distance from the reference distribution.
(b) Top: standard deviations $\sigma(\mathbb{P}_{(i)})$ of the fidelity distributions based on EM-QPT and ML-QPT for a testing dataset comprising $10^2$ QPT experiments on \texttt{X} gates.
Bottom: normalized $W_1$ distance of EM-/ML-QPT. The blue and green shaded regions indicate cases where the $W_1$ distance is smaller than that of std-QPT and RB, respectively.
Parameters: $\xi_{\rm max} = 5, \xi_{\rm min} = 1, \xi_0 = 4$.
}\label{fig:figure_last}
\end{figure}
We next evaluate the performance of our method through empirical information theory~\cite{PhysRev.106.620,majda2011improving}, which focuses on the behavior of information measures in practical, finite-sample settings.
To quantify the precision and sensitivity of gate characterization, we calculate the distance between the empirical $\mathbb{P}_{(i)}(q)$ and reference $\mathbb{P}_{\rm ref}(q)$ probability distributions by the one-dimensional Wasserstein (earth mover's) distance~\cite{panaretos2019statistical}
\beq\label{eq:W1}
W_1(\mathbb{P}_{(i)},\mathbb{P}_{\rm ref})
=\int_{\mathbb{R}}^{} |\mathcal{C}_{(i)}(q)- \mathcal{C}_{\rm ref}(q)| dq,
\eeq
where $\mathcal{C}_P(q')=\text{Prob}[q \le q']$ is the cumulative distribution function (CDF) of the probability distribution $P$.
The $W_1$ distance is the area between the CDF curve $P$ and the reference; see the lower panel in \figpanel{fig:figure_last}{a} as an example.
This distance directly captures first-moment deviations and provides an informative proxy for the second moment.

In our case, the statistical variable in Eq.~\eqref{eq:W1} is the logarithm of infidelity: $q=\log_{10}(1-\mathcal{F})\in[-\xi_{\rm max},-\xi_{\rm min}]$ with constants $\xi_{{\rm max}/{\rm min}} \in \mathbb{R}^+$.
For simplicity, we set the reference probability distribution as a delta function $ \mathbb{P}_{\rm ref} = \Delta(-\xi_{0})$ 
referring to the ideal measurement protocol that always perfectly estimates the gate fidelity $\mathcal{F}_0 = 1-10^{-\xi_0}$ fulfilling $\xi_0\in[\xi_{\rm min},\xi_{\rm max}]$.
The $W_1$ distance then obeys $W_1\in[0,\xi_{\rm max}-\xi_{\rm min}]$, where $W_1=0$ is the ideal measurement scheme that gives the exact gate fidelity. The larger the $W_1$ distance, the further away from the perfect characterization.

In \figpanels{fig:figure_last}{a}{b}, we compare the normalized $W_1/(\xi_{\rm max}-\xi_{\rm min})$ distance of the EM-QPT and ML-QPT methods
for $10^2$ QPT experiments of an \texttt{X} gate, using $10^3$ realistic error matrices and their corresponding digital twin, respectively.
We take the distribution of std-QPT and RB as delta functions, since the uncertainty for an individual QPT is negligible.
In \figpanel{fig:figure_last}{a}, the probability distribution of EM-QPT (pink) contains all types of errors in error matrices, implying large variance in fidelity estimation due to abnormal errors. 
Consequently, ML-QPT yields more reliable fidelity estimates, leading to a smaller $W_1$ distance than EM-QPT in \figpanel{fig:figure_last}{b}.

\textit{Discussion and conclusion.}
We have investigated nearly self-consistent and SPAM-error-mitigated quantum process tomography (EM-QPT) by constructing noisy probes from identity process matrices.
Moreover, we proposed machine-learning-assisted QPT (ML-QPT), further enhancing EM-QPT by fully leveraging knowledge of SPAM errors hidden in identity process matrices, enabling accurate and high-precision QPT for reliable and practical applications across a variety of quantum devices.
The SPAM-aware digital twin improves gate characterization beyond standard methods, allowing accurate fidelity estimation up to the second moment. Both numerical simulations and experimental results demonstrate that our method achieves at least an order-of-magnitude improvement in precision over standard QPT. Furthermore, we have discussed the experimental feasibility of our approach: the model demonstrates stability without time drift in practical implementations, and the training exhibits reliable convergence~\cite{SM}.
Compared to gate set tomography~\cite{robin-quantum-2021,roth-prxquantum-2023}, our method offers advantages in terms of experimental complexity and generality; see the End Matter for details. In particular, ML-QPT is more resilient to anomalous errors than other methods (see Appendix G in \cite{SM}). To further improve upon our method, one could leverage prior knowledge of SPAM errors~\cite{blumekohout2024easybetterquantumprocess} (although ML-QPT already performs well without such knowledge; see Appendix H in \cite{SM}), or advance the generative model~\cite{cGAN} to find a higher-performance digital twin of the error matrix.

A possible extension of our method is to diagnose the type of SPAM error; for example, the particular behavior of coherent and incoherent errors, providing a useful reference for experimental design. The digital twin of SPAM errors can also serve as a sensitive sensor to detect anomalies in realistic experiments~\cite{huang2024quantum}.
More broadly, our approach can be directly extended to $N$-qubit ($N > 2$) quantum processes~\cite{chris-npj-2023,mkadzik2022precision}, providing an efficient toolkit in quantum technology, e.g., for gate optimization~\cite{QOC2014RB,RL2021prxq}. Furthermore, our EM-QPT (ML-QPT) protocol has great potential in ancilla-assisted QPT, where input states are often entangled and the measurement schemes involve intricate global measurements with complex unitary operations~\cite{xue-prl-2012, patel-arxiv-2025}.

\textit{Code availability.} The codes that support the findings of this study are openly available in the repository \href{https://github.com/huangtangy/EM-QPT}{https://github.com/huangtangy/EM-QPT}.

\begin{acknowledgments}
\textit{Acknowledgments.}
This work has received funding from the EU Flagship on Quantum Technology through HORIZON-CL4-2022-QUANTUM-01-SGA Project No.~101113946 OpenSuperQPlus100. 
The Chalmers team acknowledges financial support by the Knut and Alice Wallenberg through the Wallenberg Center for Quantum Technology (WACQT).
A.F.K. is also supported by the Swedish Foundation for Strategic Research (grant numbers FFL21-0279 and FUS21-0063).
The quantum chips used in this work were fabricated at Myfab Chalmers.
The work by the Aalto researchers has been done under the Academy of Finland Centre of Excellence program (Project No. 352925).
The Aalto team acknowledges the provision of facilities and technical support by the Aalto University at the national research infrastructure OtaNano Low Temperature Laboratory.
\end{acknowledgments}

\textit{Author Contributions.} T.H.~and A.G.~conceived the idea. I.M., A.A., M.K., Y.C., O.S., and T.H.~performed the experiments and analyzed the data. T.H.~also contributed to the numerical simulations and the development of the machine learning algorithms. T.H., A.G., and I.M.~wrote the manuscript, and A.F.K., T.A., G.S.P., and G.T.~contributed to its revision. G.S.P., A.F.K., and G.T.~provided supervision and guidance throughout the project. E.H., C.W., I.A., J.Bi., A.O., M.D., M.R., A.F.R., A.N., L.C., and J.By.~participated in the device fabrication. All authors contributed to discussions and interpretation of the results.

\bibliography{ref}


\newpage
\section{End Matter}

\textit{Experimental setup.} 
Here, we detail the experimental design for varying SPAM errors. Specifically, we introduce both incoherent and coherent SPAM errors by modifying the optimized pulse envelopes in the experiments. Before proceeding, we calibrate the single-qubit gates, which rely on DRAG pulses~\cite{DRAG}, to achieve a randomized benchmarking (RB) fidelity of \(\mathcal{F}_{\rm rb} \approx \SI{99.96}{\percent} \).
In QPT experiments for an \( N \)-qubit gate, one needs to implement \( 12^N \) circuits, corresponding to the preparation of four initial states and three measurement rotations per qubit. Each circuit execution consists of five steps:
(1) Active reset of the qubits;
(2) Apply \( U_1 \) for state preparation;
(3) Perform the target gate operation \( \mathcal{E} \);
(4) Apply the rotation gate \( U_2 \);
(5) Read out all qubit states. 
Here, both \( U_1 \) and \( U_2 \) are composed of single-qubit gates. We refer the reader to the Supplemental Material~\cite{SM} for more details about experimental setups.
\begin{figure}
\centering
\includegraphics[width=0.9\columnwidth]{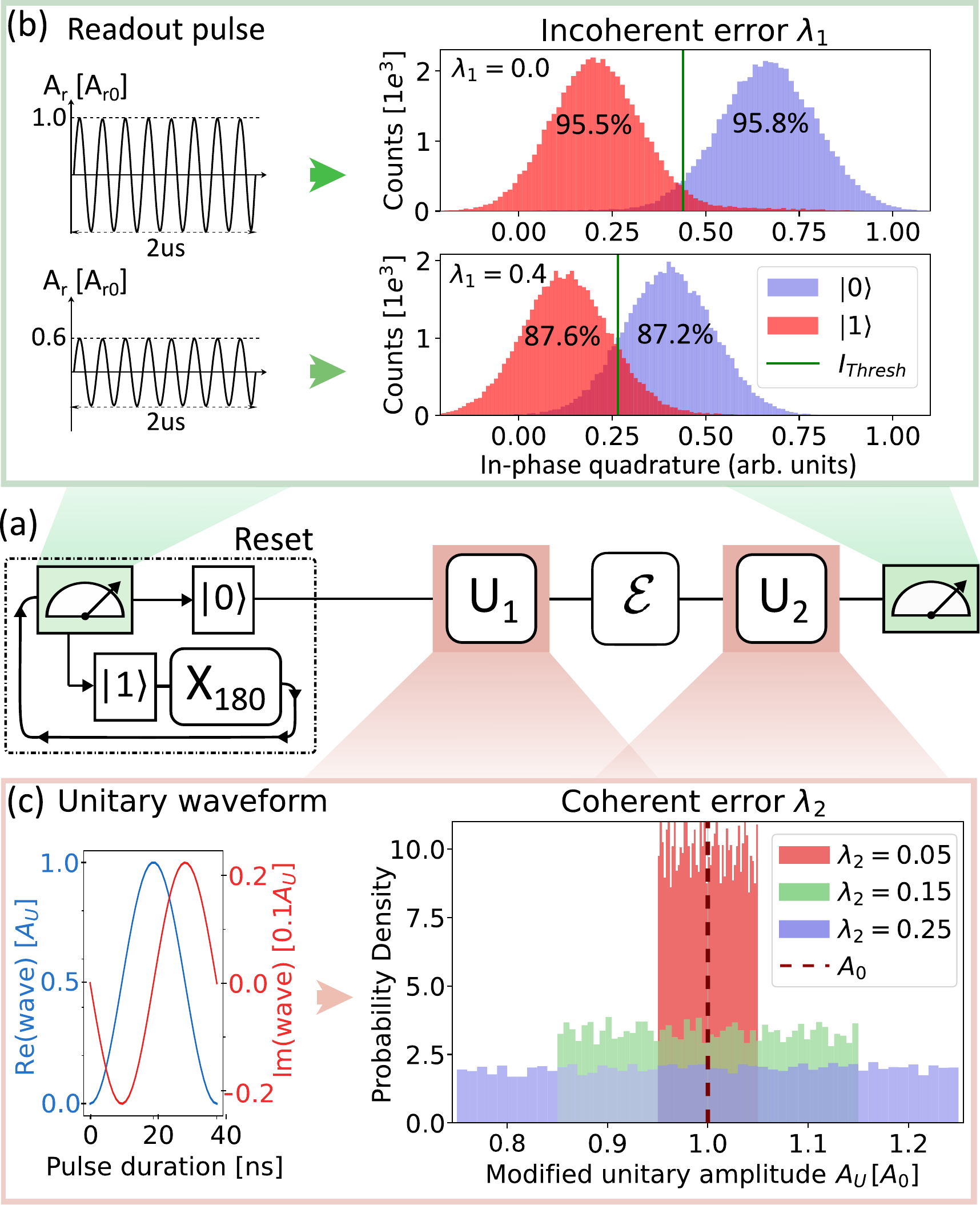}
\caption{\textbf{Schematic diagram of QPT and implementation of noise channels.} (a) Quantum process tomography with active reset (dashed rectangle) for the initialization of the qubit in the ground state $|0\rangle$. $U_1$ and $U_2$ are the sets of unitary rotations responsible for initial state preparation and measurement projectors, respectively. $\mathcal{E}$ is the process under study. 
(b) Incoherent noise channel, where the amplitude of the readout pulse (left graph) are scaled by a factor of $(1-\lambda_1)$, resulting a biased readout threshold (vertical green line in the right graph). 
Readout signal histograms visualize the separation between the peaks from the ground $|0\rangle$ (blue) and excited $|1\rangle$ (red) states for different $\lambda_1$ values. 
(c) Coherent noise channel, the amplitude $A_U$ of the unitary waveforms (left graph) is modified 
according to $A_U = A_0(1+r)$ with a uniformly sampled factor $r\sim \mathcal{U}(-\lambda_2,\lambda_2)$ (right graph). 
}
\label{fig:figure_QPT}
\end{figure}

\textit{(i) Incoherent error.} 
We introduce an additional incoherent noise channel by reducing the amplitude of the readout pulse according to the expression $A_r=A_{r0}(1-\lambda_1)$, while simultaneously scaling assignment threshold by the same factor, $(1-\lambda_1)$. Here, $A_{r0}$ denotes the optimal readout amplitude, calibrated in the absence of additional noise ($\lambda_1=0$). In each experiment, we simultaneously scale the readout pulses from the first step (active reset) and the fifth step (readout) in the same manner.
The reduced amplitude of the readout pulses leads to poor separation between the histograms corresponding to the $|0\rangle$ and $|1\rangle$ states, resulting in lower readout fidelity and less reliable ground-state initialization.
To illustrate the impact of noise, \figpanel{fig:figure_QPT}{b} presents a comparison of one-dimensional histograms at two different noise levels: $\lambda_1=0$ (top graph) and $\lambda_1=0.4$ (bottom graph). In the absence of noise, the two-state mean assignment fidelity is $F_{\text{assign}({|0\rangle,|1\rangle})} = \SI{95.65}{\percent}$ and it reduces to \SI{87.40}{\percent} for $\lambda_1=0.4$.

\textit{(ii) Coherent error.} 
We introduce a coherent noise channel by adding amplitude uncertainties to single-qubit DRAG pulses $\Omega(t)=A_0[1-\cos(2\pi t/t_g)]$~\cite{DRAG} with the gate length $t_g=\SI{40}{\nano\second}$, see \figpanel{fig:figure_QPT}{c}, left graph. 
Specifically, we use the amplitude $A_U = (1 +r) A_0$ with $r \sim \mathcal{U}(-\lambda_2, \lambda_2)$, where $A_0$ is the optimal amplitude (calibrated as described in \cite{SM}) at $\lambda_2=0$. For clarity, the probability densities of the modified unitary amplitude are presented in \figpanel{fig:figure_QPT}{c} (right graph) for different values of $\lambda_2$. In the experiment, the random variable $r$ is sampled $10^4$ times within the outer averaging loop, and the corresponding amplitude correction is applied to both $U_1$ and $U_2$.

\textit{Experimental complexity.} 
Here, we compare experimental complexity in various QPT methods: std-QPT, EM-QPT, ML-QPT, and long-sequence gate set tomography (LSGST)~\cite{robin-quantum-2021}.  Here, experimental complexity refers to the total number of experiments required to perform tomography of an unknown quantum process on a given hardware platform.

The gate sets used in QPT for state preparation and measurement are 
$\{\texttt{I}, \texttt{R}_x(-\pi/2), \texttt{R}_y(-\pi/2), \texttt{X}\}$ and $\{ \texttt{I}, \texttt{R}_x(\pi/2), \texttt{R}_y(\pi/2) \}$, respectively. Together, these form a comprehensive gate set: $\mathbb{G}=\{ \texttt{I}, \texttt{R}_x(\pm\pi/2), \texttt{R}_y(\pm\pi/2), \texttt{X} \}$, comprising six distinct gates.

Note that EM-QPT targets only the first moment (accuracy), whereas ML-QPT enhances both accuracy and precision. Their scopes of performance differ fundamentally, and a direct complexity comparison would therefore be misleading. We thus benchmark EM-QPT against std-QPT, and ML-QPT against LSGST in the following analysis.

In Table~\ref{table:complexity}, we list the experimental complexities we find. We see that EM-QPT only doubles the complexity of std-QPT for an \(N\)-qubit gate, requiring \(2 \times 12^{N}\) circuits, while achieving significantly higher accuracy.
The complexity for ML-QPT is estimated based on $N_{x}=10^2$ error matrices, along with an additional experiment for QPT of the target process, resulting in $(N_{x}+1)\times 12^{N}$ experiments for an $N$-qubit gate. For LSGST of single-qubit gate, a maximum sequence length of 16 is used, resulting in 2904 experiments~\cite{chris-npj-2023} generated using {\tt pyGSTi }~\cite{erik-qst-2020} to perform GST on the SPAM gate set $\mathbb{G}$, plus 12 additional experiments for QPT of the target process. 
For a two-qubit gate, the LSGST requires 15925 circuits when using the gate set $\{ \texttt{I}, \texttt{R}_x(\pi/2), \texttt{R}_y(\pi/2)\}^{\otimes2}$, which is generated by the predefined module ${\tt smq2Q\_XXYYII}$~\cite{erik-qst-2020}. In principle, the complete gate set $\mathbb{G}^{\otimes 2}$ should be implemented for a full two-qubit gate characterization within LSGST, which is not defined in {\tt pyGSTi}~\cite{erik-qst-2020}. Consequently, the total number of required experiments is expected to be significantly larger than the example presented here.

\begin{table}[h!]
\centering
\begin{tabular}{c @{\hspace{1cm}} c @{\hspace{0.5cm}} c}
    \hline \hline
    \textbf{Methods} & \multicolumn{2}{c}{\textbf{Experimental complexity}} \\
    \cline{2-3}
    & $N=1$ & $N=2$ \\
    \hline \hline
    std-QPT & 12 & 144 \\
    EM-QPT & 24 &  288 \\
    ML-QPT & 1212 & 14544 \\
    LSGST & 2916 & >16069 \\
    \hline
\end{tabular}
\caption{ Experimental complexity comparison for different QPT methods at $N=1$ and $N=2$.
}
\label{table:complexity}
\end{table}

We also emphasize that, in the case of ML-QPT, once the digital twin is trained, the experimental cost for performing QPT on a given N-qubit gate is only $12^N$ experimental circuits.
In contrast, in the presence of an experimental anomaly, the entire GST might fail to provide a faithful fidelity estimation since it lacks the statistical precision and robustness offered by ML-QPT on a given hardware.


\newpage

\onecolumngrid
\pagebreak
\widetext
\begin{center}
\textbf{ \large Supplemental Material of \\ “Quantum Process Tomography with Digital Twins of Error Matrices"}
\end{center}

\appendix
\onecolumngrid

For simplicity and clarity in our derivations, we use the {Pauli-Liouville representation} as a framework. The Pauli set reads
\beq\label{eq:PL_basis}
\mathcal{P} := \mleft\{ \bigotimes_{v=1}^{N}E_v : E_v \in \{I, \sigma_x, \sigma_y, \sigma_z\} \mright\},
\eeq
where $I$ is the identity matrix, and $\sigma_x, \sigma_y$, and $\sigma_z$ are the standard Pauli matrices. In this framework, the density matrix \(\rho\) and a quantum channel \(\mathcal{E}(\rho)\) can be expressed as
\beq
\rho = \sum_{j=1}^{4^N} a_j E_j, \quad \mathcal{E}(\rho) = \sum^{4^N}_{\alpha,\beta=1} \chi_{\alpha\beta} E_\alpha \rho_0 E_\beta^\dagger,
\eeq
where \(E_j \in \mathcal{P}\) and \(a_j = \text{Tr}(E_j^\dagger \rho)/{2^N}\). To facilitate further analysis, we vectorize the density operator:
\beq
\text{vec}(\rho) = |\rho\rrangle = (\rho_1, \rho_2, \dots, \rho_{4^n})^{T}, \quad \mathcal{E}(\rho) = \chi \circ |\rho\rrangle.
\eeq
Here, the symbol $A \circ B$ denotes the \textit{completely positive trace-preserving} (CPTP) mapping of $A$ acting on $B$. Additionally, we note the trace property
\beq
\text{Tr}[\hat{O}\rho] = \text{vec}(\hat{O}^T) |\rho\rrangle = \llangle \hat{O} | \rho \rrangle.
\eeq

\subsection{Appendix A: Self-inconsistency in standard quantum process tomography}

In quantum process tomography (QPT), the objective is to reconstruct the quantum gate operation $\mathcal{E}_{\rm g}$ using a set of well-defined probes, which include a collection of input states $\{\rho_i\}$ and measurement operators $\{M_\mu\}$. 
In the ideal scenario, the measurement outcomes are given by  
\beq
p_{i,\mu} = \llangle M_{\mu} | \mathcal{E}_{\rm g} | \rho_i \rrangle,
\eeq
where the process $\mathcal{E}_{\rm g}$ can be accurately determined from QPT experiments.
In practice, however, imperfections in state preparation and measurement (SPAM) introduce systematic errors, leading to noisy measurement outcomes (we use a tilde to denote variables affected by noise such that they give faulty values):  
\beq
\tilde{p}_{i,\mu} = \llangle M_{\mu} | \mathcal{E}_{\rm m}^{\dagger} \circ \mathcal{E}_{\rm g} \circ \mathcal{E}_{\rm sp} | \rho_i \rrangle,
\eeq
where SPAM errors are modeled as two distinct error channels, $\mathcal{E}_{\rm sp}$ and $\mathcal{E}_{\rm m}$, acting on the ideal states and measurements:
\beq
\tilde{\rho}_{i} = \mathcal{E}_{\rm sp} | \rho_i \rrangle, \quad 
\tilde{M}_{\mu} = \mathcal{E}_{\rm m} | M_\mu \rrangle.
\eeq

In the absence of SPAM errors, the estimated process matrix $\chi_{\rm e}$ is calculated as
\beq\label{eq:chi_ML}
\mathcal{J}(\rho_i, M_{\mu}, p_{i,\mu}) \rightarrow \chi_{\rm e} \quad ({\text{ideal case}}),
\eeq
using some data post-processing QPT algorithm $\mathcal{J}$.
In practice, this ideal scenario is never realized due to unavoidable imperfections in the SPAM steps, resulting in noisy data $\{\tilde{p}_{i, \mu}\}$, and leading to an incorrect estimation of process matrix:
$\mathcal{J}(\cdot, \cdot, \tilde{p}_{i,\mu}) \rightarrow \tilde{\chi}_{\rm e}$.
This noisy case can be further categorized into three distinct scenarios, depending on the extent of knowledge about the SPAM errors:
\begin{align}
\mathcal{J}(\tilde{\rho}_i, \tilde{M}_{\mu}, \tilde{p}_{i,\mu}) \rightarrow \tilde{\chi}^1_{\rm e} \quad ({\text{self-consistent}}), \label{case1} \\
\mathcal{J}(\rho_i, M_{\mu}, \tilde{p}_{i,\mu}) \rightarrow \tilde{\chi}^2_{\rm e} \quad ({\text{self-inconsistent}}), \label{case2} \\
\mathcal{J}(\bar{\rho}_i, \bar{M}_{\mu}, \tilde{p}_{i,\mu}) \rightarrow \tilde{\chi}^3_{\rm e} \quad ({\text{$\approx$ self-consistent}}). \label{case3}
\end{align}
In the self-inconsistent case [Eq.~\eqref{case2}], the estimated process matrix includes both  gate operation and SPAM errors:
\beq
\tilde{\chi}_e^2: \tilde{\mathcal{E}}= \mathcal{E}_{\rm m} \circ \mathcal{E}_{\rm g} \circ \mathcal{E}_{\rm sp}.
\eeq
Therefore, standard QPT (std-QPT) yields a biased estimate of the true gate, as it assumes perfect probes while relying on noisy measurement data. Consequently, one cannot distinguish whether the observed process originates from the gate itself or from the SPAM. 
The notion of \textit{self-inconsistency} thus refers to reconstructing a quantum process from noisy experimental data while presuming ideal measurement probes. On the other hand, we interpret \textit{self-consistency} as the ability to reconstruct the gate operation from experimental data with explicit knowledge of SPAM errors, enabling accurate and unbiased characterization even in their presence.

\subsection{Appendix B: Estimation of noisy probes}

Equation~\eqref{case3} represents an implementable yet nearly self-consistent version of QPT~\cite{blumekohout2024easybetterquantumprocess}.
Here, $\{\bar{\rho}_i, \bar{M}_{\mu}\}$ are experimentally tomographed noisy probes, obtained with certain assumptions and approximations, yielding $\tilde{\chi}^3_{\rm e} \approx \tilde{\chi}^1_{\rm e}$. Due to gauge symmetry, it is inherently impossible to tomograph states and POVMs independently with arbitrary accuracy and precision, as quantum state tomography (QST) relies on POVMs as probes and vice versa~\cite{laflamme-prr-2021, robin-natcom-2017, karol-quantum-2018, lin-njp-2019}. 
For example, the state (POVM) tomography of an unknown state $\rho$ admits the property of gauge symmetry (unitary invariance):
\[p_\mu = \llangle {M}_\mu|{\rho }\rrangle\equiv\llangle {M}_\mu| U^{\dagger}U|{\rho}\rrangle,\]
under unitary transformation $U$. In other words, one cannot discriminate whether the readout results comes from $\{|M_\mu\rrangle\}$ or $\{U|M_\mu\rrangle\}$, so the state cannot be uniquely reconstructed. In this vein, quantum state tomography relies on given POVMs and vice versa. For practical applications, this case emerges as the most promising and reliable alternative to the self-inconsistent case, Eq~\eqref{case2}, as it effectively mitigates the SPAM errors in QPT and enables faithful reconstruction in a self-consistent manner.

The objective of this section is to estimate the noisy probes, including both quantum states and measurements, based on the process tomography of the idling process. Specifically, we aim to perform state and detector tomography by leveraging the error matrix. To achieve this, we define a loss function that incorporates the state \(\rho\), measurement \(M\), and process matrix \(\chi\):
\beq\label{loss_J}
\mathcal{J}(M, \rho, \chi) = \big\|\textbf{A} |\chi\rrangle - \llangle M | \rho \rrangle \big\|,
\eeq
where $\parallel \cdot \parallel$ represents the $L_2$ norm and $\textbf{A}$ is the \textit{sensing matrix}, which can be computed analytically~\cite{korotkov-prb-2014}. 
The loss function in Eq.~\eqref{loss_J} quantifies the discrepancy between the ideal readout and the experimental outcomes obtained for given initial states and measurement setups. 

Given the process matrix \(\tilde{\chi}^{I}\) obtained from identity process tomography, the noisy initial states can be estimated by solving the optimization problem
\begin{align}
&\overline{\rho}_i = {\argmin_{\rho_i}} \sum_\mu \mathcal{J}(\rho_i, M_\mu,  \tilde{\chi}^{I}) \nonumber \\
\quad \text{subject to} &\quad \text{Tr}(\rho_i) = 1, \quad \rho_i^\dagger = \rho_i, \quad \rho_i \ge 0.
\end{align}
Here, we assume that the measurement probes \(\{M_\mu\}\) are perfect, and the optimized states satisfy the trace condition $\text{Tr}(\rho_i) = 1$, Hermiticity ($\rho^\dagger = \rho_i$), and positive semi-definiteness ($\rho_i \ge 0$). 

Similarly, the noisy measurement probes can be estimated by
\beqa
\{\overline{M}_\mu\} = \argmin_{\{ {M}_\mu\}} \sum_{i} \mathcal{J}(\rho_i, \{ {M}_\mu\}, \tilde{\chi}^{I}) \nonumber \\
\quad \text{subject to} \quad \sum_\mu M_\mu = \mathbb{I}, \quad M_\mu \ge 0,
\eeqa
where the initial states are assumed to be perfect, and the estimated positive operator-valued measures (POVMs) satisfy the properties of positivity and completeness. These optimizations are performed using the Python-based library \texttt{cvxpy}~\cite{diamond2016cvxpy}.

\subsection{Appendix C: EM-QPT on an arbitrary $N$-qubit quantum channel}
In this section, we investigate the performance of our method for a general CPTP quantum channel, which can be expressed in the Pauli basis.
The Kraus operators $\{\mathcal{K}^{(i)}\}$ can be expressed in the Pauli basis $\{E_k\}$ [Eq.~\eqref{eq:PL_basis}] as
\beq
\mathcal{K}^{(i)}=\sum_k^{4^N} c^{(i)}_k E_k ,
\eeq
where $c_k^{(i)} \in \mathbb{C}$ are complex coefficients. For an $N$-qubit system, any CPTP map can be constructed using up to $r \leq 4^N$ Kraus operators:
\beq
\mathcal{E}_{\text{rand}}(\rho)= \sum^{r\le 4^N}_{i=1} \mathcal{K}^{(i)}\rho(\mathcal{K}^{(i)})^{\dagger},
\eeq
with the trace-preserving condition requiring $\sum_{i=1}^r (\mathcal{K}^{(i)})^\dagger \mathcal{K}^{(i)} = \mathbb{I}$.

To numerically generate valid CPTP maps, we first sample random complex coefficients $\{c_k^{(i)}\} \in \mathbb{C}^{4^N \times r}$ to form unnormalized Kraus operators $\{\mathcal{K'}^{(i)} = \sum_k c_k^{(i)} E_k\}_{i=1}^r$, and then normalize them to satisfy the TP condition: 
\beq
\mathcal{K}^{(i)} = \frac{\mathcal{K'}^{(i)}}{\sqrt{\sum^r_{i=1} \mathcal{K'}^{(i)} (\mathcal{K'}^{(i)})^{\dagger}} }.
\eeq
The resulting process matrix $\chi$, defined through  
\beq
\chi_{\alpha\beta} = \sum_ic^{(i)}_\alpha (c^{(i)}_\beta)^*,
\eeq
is guaranteed to be Hermitian and completely positive by construction.

\begin{figure}
\centering
\includegraphics[width=0.5\columnwidth]{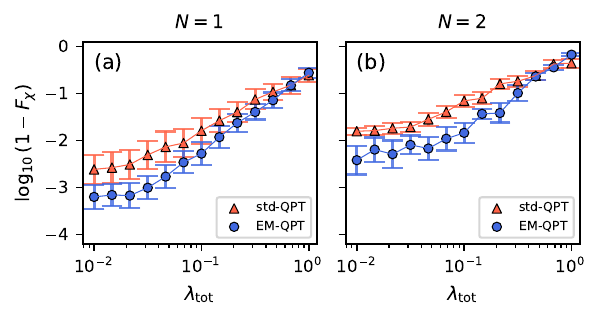} 
\caption{\textbf{EM-QPT of general CPTP processes for one- and two-qubit quantum process.} Process infidelity as a function of total SPAM error for (a) single- and (b) two-qubit CPTP processes with randomly sampled Kraus number. The orange triangles and blue circles are the results averaged over $10^2$ random CPTP processes with an error bar of one standard deviation.
}
\label{fig:SM_cptp}
\end{figure}

We insert the quantum operation $\mathcal{E}_{\text{rand}}(\cdot)$ in the quantum circuit by using the function ${\tt Kraus}(\{\mathcal{K}^{(i)}\})$ in ${\tt qiskit }$ platform. To investigate the EM-QPT for a general CPTP process, we randomly sample the Kraus number $r\in [1,4^N]$ for $N$-qubit quantum channel. The process fidelity as a function of total SPAM error are presented in \figref{fig:SM_cptp} for one- and two-qubit channels, respectively.

\subsection{Appendix D: Digital twinning of the error matrix}

\subsubsection{1. The generative model: variational autoencoder}

A variational autoencoder (VAE)~\cite{VAE0} is a generative model that combines deep learning with probabilistic frameworks to learn a latent representation of data. Unlike traditional autoencoders~\cite{Goodfellow-et-al-2016}, a VAE models the latent space as a probability distribution rather than fixed embeddings. Specifically, a VAE consists of an \textit{encoder} that maps the input \(\textbf{x}\) to a latent distribution \(\mathbb{Q}(\textbf{z}|\textbf{x}) \sim \mathcal{N}(\mu, \sigma^2)\), where \(\mu\) and \(\sigma^2\) are learned through separate fully connected layers. The \textit{decoder} reconstructs the input data \(\textbf{x}'\) from a sampled latent vector \(\textbf{z}\). 

To enable backpropagation through the network, the latent variable \(\textbf{z}\) is sampled using the reparameterization trick~\cite{VAE1}:
\beq
\textbf{z} = \mu(\textbf{x}) + \epsilon \cdot \sigma(\textbf{x}),
\eeq
where \(\epsilon \sim \mathcal{N}(0, 1)\) is an auxiliary noise variable sampled from a standard normal distribution. 
To train the VAE model, we define the loss function, which is to be minimized during training process:
\beq
\textbf{Loss}= \mathcal{L}_{\text{recon}}(\textbf{x}',\textbf{x}) + \beta D_{\text{KL}}\left[\mathbb{Q}(\textbf{z}|\textbf{x}) \parallel \mathcal{N}(\textbf{z}|\textbf{x}) \right].
\eeq
The first term,  $\mathcal{L}_{\text{recon}}=\sum_{n,m} (\textbf{x}_{n,m}-\textbf{x}'_{n,m})^2$, is the reconstruction loss, measuring the geometric distance between  reconstructed ($\textbf{x}'$) and training $(\textbf{x})$ data.
The second term measures how different the learned latent distribution $\mathbb{Q}(\textbf{z}|\textbf{x})$ is from the normal distribution $\mathcal{N}(\textbf{z}|\textbf{x})$, quantified by the Kullback–Leibler divergence $D_{\text{KL}}\left[\mathbb{Q}(\textbf{z}|\textbf{x}) \parallel \mathcal{N}(\textbf{z}|\textbf{x})\right]$. The total loss is a combination of these two terms with a weight coefficient $\beta$.
Since both the learned posterior and the target prior are Gaussian distributions, the KL divergence has a closed-form solution~\cite{VAE0}:
\beq
D_{\text{KL}}\mleft[\mathbb{Q}(\textbf{z}|\textbf{x}) \parallel \mathcal{N}(0, I)\mright] = \frac{1}{2} \mleft( \sigma^2 + \mu^2 - 1 - \log \sigma^2 \mright).
\eeq

\subsubsection{2. Unsupervised learning of error matrices}
The goal of this section is to learn the state preparation and measurement (SPAM) errors by training a generative model on realistic experimental data. We hypothesize that a deep neural network (NN)-based model can statistically generate the idle process matrix, capturing the features of SPAM errors. This NN serves as a \textit{digital twin}~\cite{huang2024quantum} of the SPAM error.

To achieve this, we first construct a training database \(\textbf{X} = \{\textbf{x}^{(i)}\}_{i=0}^{N_x}\), consisting of \(N_x\) independent quantum process tomography (QPT) experiments for the idling quantum process. For an \(N\)-qubit quantum channel, the error matrix is a complex matrix \(\chi_i \in \mathbb{C}^{4^{N} \times 4^{N}}\). Each \(\chi\) matrix is decomposed into real and imaginary parts, forming the input training data points \(\textbf{x}_i \in \mathbb{R}^{2 \times 4^{N} \times 4^{N}}\).

In this framework, the digital twin of the error matrix, \(\textbf{x}'_i \to \chi'_i\), is expected to statistically mitigate SPAM errors for an arbitrary quantum process $\mathcal{E}$. The neural network learns the underlying features of SPAM errors embedded in the error matrix, enabling enhanced QPT with the assistance of the digital twin.

The \textit{encoder} extracts key features from the input data \(\textbf{x}_i\) using a series of convolutional layers~\cite{Goodfellow-et-al-2016}, progressively reducing the spatial dimensions while increasing the number of channels. At the end of this process, the high-dimensional input is transformed into a low-dimensional feature representation. Instead of encoding the input into a fixed lower-dimensional representation \(\textbf{x} \to \textbf{z}\), the VAE learns a probabilistic distribution characterized by two variational parameters \(\{\mu, \sigma^2\}\), such that \(\textbf{z} \sim \mathcal{N}(\mu, \sigma^2)\).

To ensure that the VAE output satisfies the CPTP constraint, we introduce an additional layer called {\tt QProcess}, which relies on the Cholesky decomposition~\cite{PhysRevA.61.010304}$: \mathbf{B} = \mathbf{L} \mathbf{L}^T$, where $\mathbf{L}$ is a lower triangular matrix and $\mathbf{L}^T$ is its transpose. Cholesky decomposition is a powerful mathematical tool that plays a significant role in quantum information science, both in QPT~\cite{Shahnawaz2021cGAN} and quantum state tomography (QST)~\cite{Gaikwad2025}.
To ensure numerical stability, the transformed output is $\mathbf{x} = \mathbf{L} \mathbf{L}^T + \epsilon \mathbf{I}$,
where $\epsilon$ is a small positive number (e.g., $10^{-5}$).

As a result, the forward data flow of the model is
\beq
\textbf{x} \to \tt{Encoder}  \to \textbf{z} \to \tt{Decoder}\to \tt{QProcess} \to \textbf{x}'.
\eeq
For a well-trained model (denoted $^*$), there is a straightforward approach to evaluating the VAE model. As mentioned earlier, the goal of VAE is to map the training data set to a Gaussian distribution that closely approximates the standard normal distribution. In this context, one can verify the distribution of the latent vectors by passing the training data through the trained encoder as
\beq
\textbf{x} \to \tt{Encoder}^* \to \textbf{z}',
\eeq
where, ideally, \( \textbf{z}'\sim\mathcal{N}(0,\mathbb{I}) \).

\begin{figure}
\centering
\includegraphics[width=0.6\columnwidth]{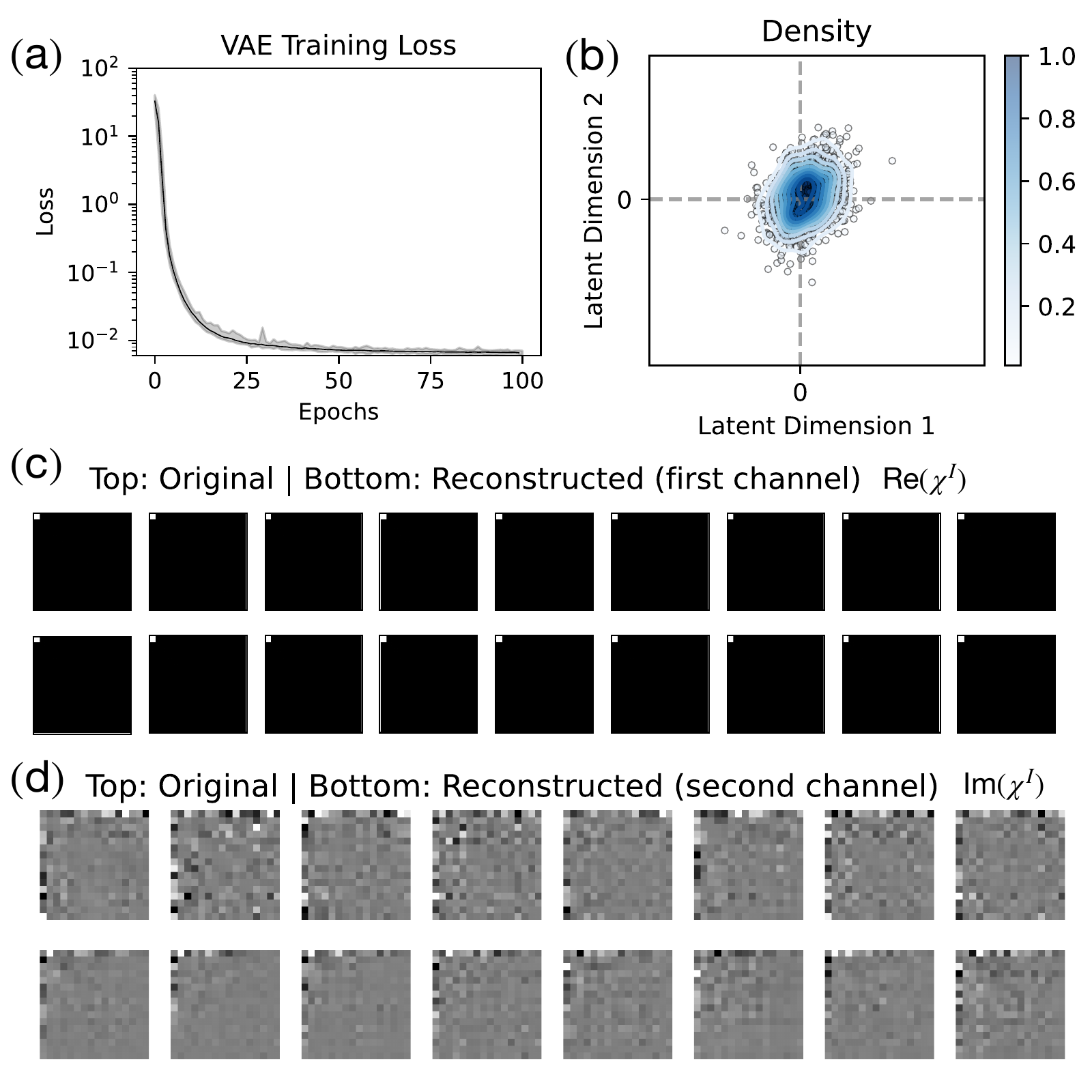}
\caption{\textbf{Training outcomes of the VAE model for generating digital twins of two-qubit error matrices.} 
(a) Evolution of the training loss over 100 epochs, showing a rapid initial decrease and eventual convergence, indicating successful training. 
(b) Distribution of the latent vectors $\mathbf{z}'$ obtained by encoding the training dataset using the trained encoder. The probability density function (PDF) is estimated via Gaussian kernel density estimation (blue curves), demonstrating that the learned latent representation closely follows a standard normal distribution. 
(c) Comparison between original (top) and reconstructed (bottom) error matrices for the real component (first channel) of the two-qubit gate errors. The reconstructed matrices are generated by decoding latent vectors sampled from the trained encoder. 
(d) Similar comparison for the imaginary component (second channel) of the error matrices. 
The results indicate that the VAE successfully captures the statistical properties of the error matrices, enabling the generation of realistic digital twins for quantum process tomography and error mitigation. The total error rate is $\lambda_{\text{tot}}=0.01$ and the size of traning data sets is $10^3$.
}
\label{fig:figure_SM2}
\end{figure}

In \figpanels{fig:figure_SM2}{a}{b}, we illustrate the evolution of the loss during training and the distribution of the latent vectors \( \textbf{z}' \) across the entire training dataset. The probability density function (PDF) of \( \textbf{z}' \) is estimated using Gaussian kernel density estimation~\cite{bashtannyk2001bandwidth} and is depicted by the blue curves in \figpanel{fig:figure_SM2}{b}. The results indicate that the two-dimensional latent vectors \( \textbf{z}' \) produced by the trained encoder are close to a standard normal distribution.

Following this, digital twins can be automatically generated by sampling the latent space from a normal distribution:
\beq
\textbf{z} \sim \mathcal{N}(0,\mathbb{I}) \to \tt{Decoder}^* \to \tt{QProcess}^* \to \textbf{x}^*,
\eeq
where $\textbf{x}^*$ denotes the digital twin of error matrices.
We now present the VAE-generated digital twin of error matrices for two-qubit unitary gates with an error rate of \( \lambda_{\text{tot}} = 0.01 \). In \figpanels{fig:figure_SM2}{c}{d}, we compare the reconstructed data for the real (first channel) and imaginary components (second channel) of the error matrices with those sampled from the training dataset.
This approach enables the generation of error matrices that statistically capture the characteristics of SPAM errors, providing an efficient tool for constructing the error-mitigated quantum process tomography. 
The whole training process is performed using the \texttt{PyTorch}~\cite{pytorch} framework.
 
\subsection{Appendix E: Experimental details}
\begin{table}
\centering
\begin{tabular}{llc}
\hline
\qquad \qquad \qquad \qquad \textbf{Parameter} & & \textbf{Value} \\
\hline
Qubit frequency, & $\omega_0/2\pi$ & \SI{3.772}{\giga\hertz} \\
Qubit anharmonicity, & $\alpha_0/2\pi$ & \SI{-222}{\mega\hertz} \\
Resonator frequency, & $\omega_r/2\pi$ & \SI{6.439}{\giga\hertz} \\
Dispersive shift, & $\chi_r/2\pi$ & \SI{0.350}{\mega\hertz} \\
Relaxation time, & $T_1$ & \SI{131}{\micro\second} \\
Decoherence time, & $T_2^*$ & \SI{64}{\micro\second} \\
Spin Echo, & $T_2^{Echo}$ & \SI{141}{\micro\second} \\
Single-qubit gate fidelity, & $F_{1Q}$ & \SI{99.963}{\percent} \\
Two-state assignment fidelity, & $F_{\text{Assign}(|0\rangle,|1\rangle)}$ & \SI{95.65}{\percent} \\
\hline
\end{tabular}
\caption{\textbf{Device A.} Measured qubit parameters, coherence properties, and single-qubit performance.}
\label{tab:device_parameters}
\end{table}

\subsubsection{1. Device details}

\begin{figure}
\centering
\includegraphics[width=0.45\columnwidth]{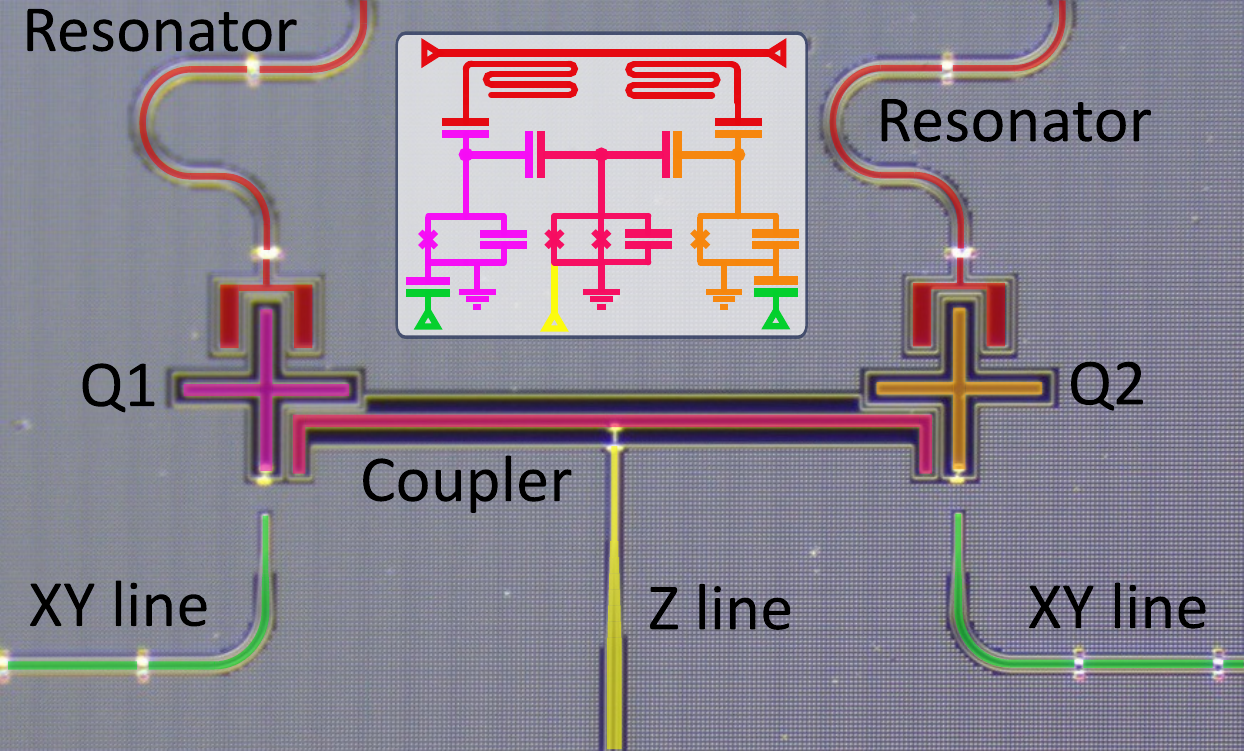}
\caption{\textbf{Single-qubit gate device micrograph with false colors.} The device used for single-qubit gate experiment consists of two fixed-frequency xmon-style transmon qubits coupled via a frequency-tunable anharmonic oscillator with SQUID. The insertion shows the circuit schematic. Only the left qubit (Q1) is used for the experiment.}
\label{fig:figure_single_qubit_gate_device}
\end{figure}

The parameters of the sample (Device A) used in the experiment with single-qubit gates are listed in Table~\ref{tab:device_parameters}. 
Device A comprises two fixed-frequency xmon-style transmon qubits~\cite{Koch2007} capacitively coupled via a frequency-tunable anharmonic oscillator. Details on the experimental setup can be found in Ref.~\cite{kuzmanovic2025neural}. Device schematic and micrograph are presented in \figref{fig:figure_single_qubit_gate_device}.
The two qubits are capacitively coupled to individual control lines and quarter-wavelength resonators for readout. The tunability of the coupler is provided by two Josephson junctions in a superconducting quantum interference device (SQUID) configuration. Our experiments were performed at the zero-flux sweet spot, where the coupler had the maximal frequency, to suppress the interaction with the qubit under investigation.

\begin{figure}
\centering
\includegraphics[width=0.6\columnwidth]{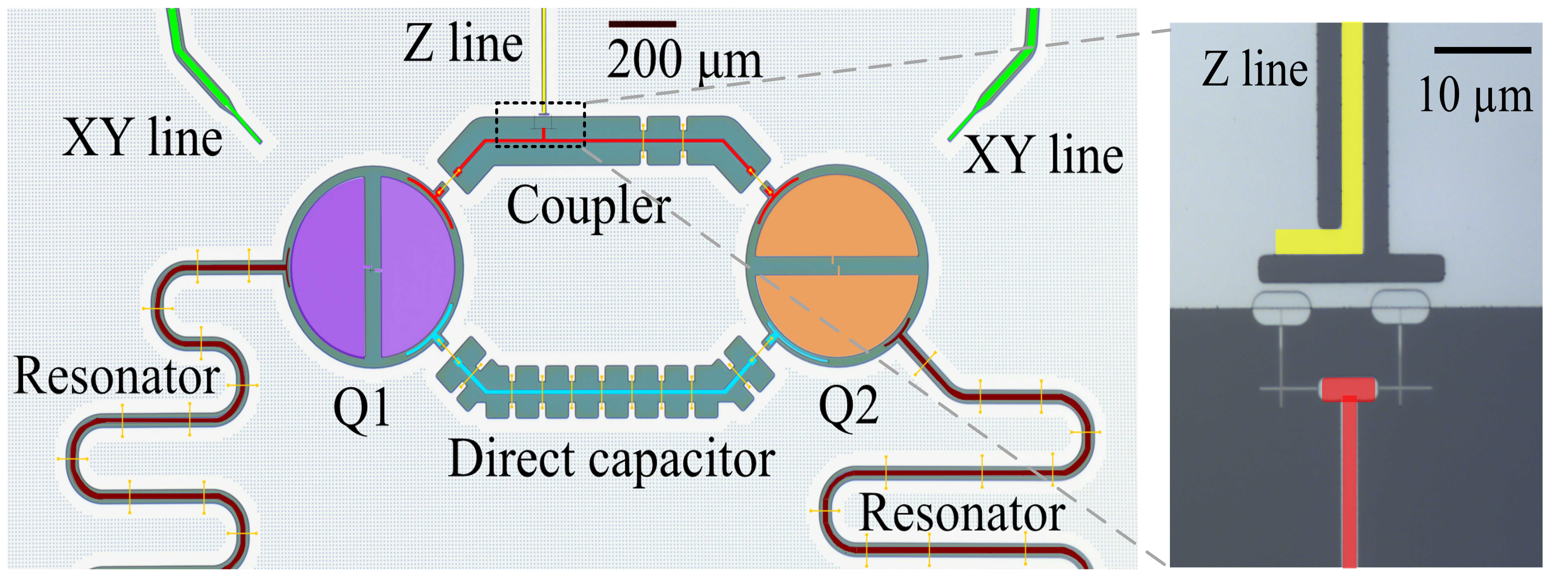}
\caption{\textbf{Two-qubit gate device micrograph with false coloring}. The device used for two qubit gate consists of two fixed-frequency transmons and one flux tunable coupler. The right-hand side image shows a zoomed-in view of the Z-line (yellow), shorted to ground, inductively coupled to the SQUID loop of the coupler.} 
\label{fig:2qb_device}
\end{figure}

To implement the two-qubit CZ gate, we used a device (Device B) with two fixed frequency transmons (Q1 and Q2) coupled via flux-tunable coupler as shown in \figref{fig:2qb_device}. The device parameters are given in Table~\ref{tab:device_parameters_2qb}. We executed the gate using baseband pulses on the coupler leading to adiabatically acquiring $\pi$ phase on the target qubit when control qubit is in excited state. 

\begin{table}[h]
\centering
\begin{tabular}{llc}
\hline
\textbf{Parameter} & \textbf{Q1} & \textbf{Q2} \\
\hline
Qubit frequency ($\omega_0/2\pi$) & \SI{4.7744}{\giga\hertz} & \SI{4.6841}{\giga\hertz} \\
Qubit anharmonicity ($\alpha_0/2\pi$) & \SI{-239}{\mega\hertz} & \SI{-239}{\mega\hertz} \\
Resonator frequency ($\omega_r/2\pi$) & \SI{6.997}{\giga\hertz} & \SI{6.852}{\giga\hertz} \\
Dispersive shift ($\chi_r/2\pi$) & \SI{0.420}{\mega\hertz} & \SI{0.390}{\mega\hertz} \\
Relaxation time ($T_1$) & \SI{66}{\micro\second} & \SI{84}{\micro\second} \\
Decoherence time ($T_2^*$) & \SI{45}{\micro\second} & \SI{57}{\micro\second} \\
Single-qubit gate fidelity ($F_{1Q}$) & \SI{99.948}{\percent} & \SI{99.966}{\percent} \\
Three-state assignment fidelity ($F_{\text{Assign}(|0\rangle,|1\rangle, |2\rangle)}$) & \SI{93}{\percent} & \SI{94.3}{\percent} \\
\hline
\end{tabular}
\caption{\textbf{Device B.} Measured qubit parameters, coherence properties, and gate performances for experiments on a two-qubit gate device.}
\label{tab:device_parameters_2qb}
\end{table}

\subsubsection{2. Experimental setups}
Experiments with single-qubit gates were conducted in a Bluefors XLD400 dilution refrigerator maintaining a base temperature stabilized at $\sim\SI{15}{\milli\kelvin}$.
The room-temperature setup, together with the full wiring diagram in our dilution refrigerator, is the same as in Ref.~\cite{kuzmanovic2025neural}. 
The drive pulses for the qubits and resonators are generated by a combination of Quantum Machines Operator-X (OPX+) quantum controller and Quantum Machines Octave. The IF (intermediate frequency) signals from the OPX+'s arbitrary-waveform generators are fed to the built-in IQ mixers of the Octave and combined with signals from the internal local oscillators. Then the outputs of the Octave are fed into qubit drive lines, which are capacitively coupled to the transmon circuits, or into the readout feedline capacitively coupled to the resonators on chip.

The transmitted readout signal propagates through the sample to a double junction circulator (\qtyrange{4}{8}{\giga\hertz} from Low Noise Factory), a directional coupler (\qtyrange{4}{12}{\giga\hertz}), and a traveling wave parametric amplifier (TWPA) fabricated at VTT Technical Research Center of Finland. After TWPA amplification the signal transmits through a single junction circulator (\qtyrange{4}{8}{\giga\hertz} GHz from QuinStar Technology), a series of low-pass, infrared, and high-pass filters, as well as an additional double junction circulator (\qtyrange{4}{8}{\giga\hertz} GHz from Low Noise Factory), before being amplified by around \SI{40}{\decibel} at \SI{4}{\kelvin} with a high-electron-mobility transistor (HEMT) from Low Noise Factory. In addition, the room-temperature microwave amplifier from Narda-MITEQ is connected to the readout chain outside of the dilution refrigerator. 
The pump signal and the DC bias current for the TWPA are provided by an Anritsu MG3692 signal generator and Yokogawa GS200, through a current limiting resistor. Circulators in the output line are essential to prevent strong signals from reflecting back into the sample. Finally, the output signal is down-converted inside the Octave to the hundreds of MHz range to be recorded and integrated by the OPX+ internal digitizer.

The schematic diagram of the measurement setup for the two-qubit gate experiment is the same as in Ref.~\cite{Anuj2025mitigating}. The device is mounted at the \SI{10}{\milli\kelvin} stage of a Bluefors LD250 dilution refrigerator. The drive pulses for the qubits and resonators are generated by a QBLOX cluster, which consists of QCM-RF, QRM-RF, and QCM modules. QCM-RF controls the single-qubit gates via the XY drive lines (see \figref{fig:2qb_device}).
Multiplexed readout is performed with QRM-RF. Flux pulses to the coupler are applied to the Z line (see \figref{fig:2qb_device}) from the QCM control module that can send arbitrary pulses from DC to \SI{400}{\mega\hertz}.

\subsubsection{3. Gate calibration} \label{A_c2:Gate_calibration}

Calibrations of the fundamental transition frequencies and coherence times provided in Table~\ref{tab:device_parameters} mainly follow well-established parameter-estimation methods used for superconducting quantum computing~\cite{Chen2018}.
Once the qubit frequency and coherence times are determined, we calibrate the amplitudes of $\pi/2$ ($\pm \text{X90}$ and $\pm \text{Y90}$) and $\pi$-pulses (\text{X180} and \text{Y180}) for later use in standard quantum process tomography (std-QPT) and randomized benchmarking (RB). 
Our single-qubit pulses use a cosine envelope with a \SI{40}{\nano\second} duration, combined with the derivative removal by adiabatic gate (DRAG)~\cite{DRAG} technique. 

The goal of the calibration is to determine the correct amplitudes of the $\pi/2$ and $\pi$ pulses, and then the corresponding DRAG coefficients. Amplitude tune up is performed via repeatedly executing pseudo-identity pulse sequences (such as two $\pi$-pulses or four $\pi/2$-pulses) $N$ times and measuring the state of the qubit across different pulse amplitudes and number of pulses. For the DRAG calibration we utilized the ``Google method''~\cite{PhysRevLett.116.020501}. 
Both protocols are implemented on the Quantum Machines OPX+ and Octave based on open-source libraries for pulse-level control of quantum bits built with the QUA programming language. The programming code is similar to the example described in 
\href{https://github.com/qua-platform/qua-libs/tree/main/Quantum-Control-Applications/Superconducting/Single-Flux-Tunable-Transmon/Use%20Case%202%20-%20DRAG%20coefficient%20calibration#1}{https://github.com/qua-platform/qua-libs}.

\subsubsection{4. Interleaved randomized benchmarking}

The performance of the single-qubit gates is characterized with Clifford RB~\cite{RB2011prl}.
To implement this technique, we generate random sequences of single-qubit Clifford gates from Table~\ref{tab:rotations} and measure the state of the qubit afterwards. Each random sequence ends with the recovery gate that will bring the qubit back to its ground state. It should be noted here that the average number of single-qubit gates in Clifford operation is 1.875.

\begin{table} 
\centering
\begin{tabular}{|c|l|}
\hline
\textbf{Index} & \textbf{Combinations} \\
\hline
1 & [\text{I}] \\
2 & [\text{X180}] \\
3 & [\text{Y180}] \\
4 & [\text{Y180}, \text{X180}] \\
5 & [\text{X90}, \text{Y90}] \\
6 & [\text{X90}, -\text{Y90}] \\
7 & [-\text{X90}, \text{Y90}] \\
8 & [-\text{X90}, -\text{Y90}] \\
9 & [\text{Y90}, \text{X90}] \\
10 & [\text{Y90}, -\text{X90}] \\
11 & [-\text{Y90}, \text{X90}] \\
12 & [-\text{Y90}, -\text{X90}] \\
13 & [\text{X90}] \\
14 & [-\text{X90}] \\
15 & [\text{Y90}] \\
16 & [-\text{Y90}] \\
17 & [-\text{X90}, \text{Y90}, \text{X90}] \\
18 & [-\text{X90}, -\text{Y90}, \text{X90}] \\
19 & [\text{X180}, \text{Y90}] \\
20 & [\text{X180}, -\text{Y90}] \\
21 & [\text{Y180}, \text{X90}] \\
22 & [\text{Y180}, -\text{X90}] \\
23 & [\text{X90}, \text{Y90}, \text{X90}] \\
24 & [-\text{X90}, \text{Y90}, -\text{X90}] \\
\hline
\end{tabular}
\caption{Single-qubit Clifford gates.}
\label{tab:rotations}
\end{table}

\begin{figure}
\centering
\includegraphics[width=0.5\columnwidth]{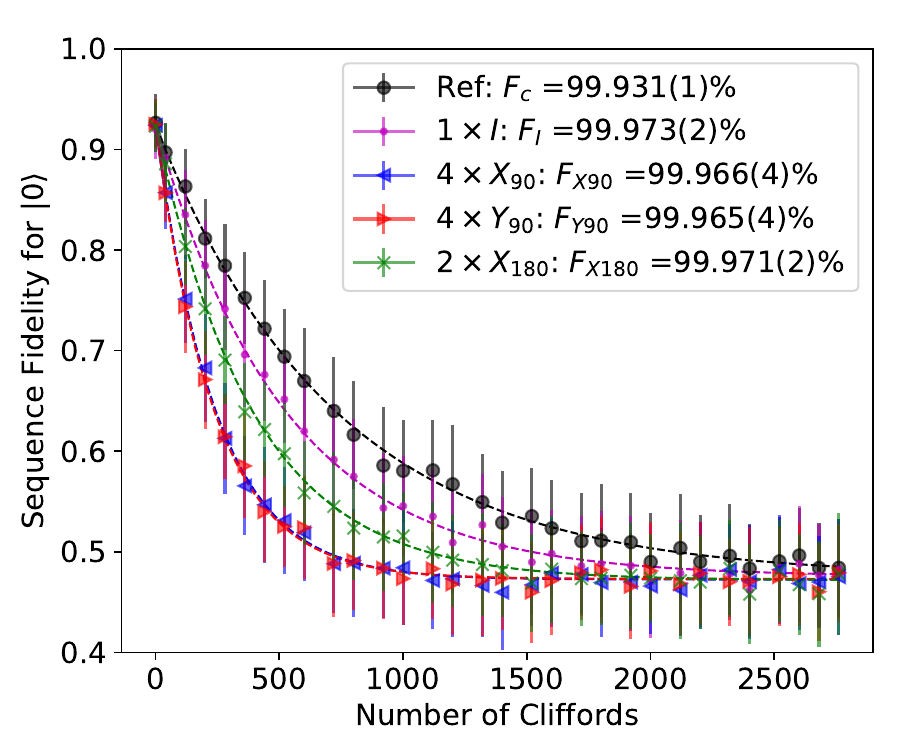}
\caption{\textbf{Interleaved randomized benchmarking (RB) of single-qubit gates.} All curves are averaged over 50 random seeds.
}
\label{fig:figure_fidelity}
\end{figure}

In the case of the interleaved RB, we are aiming to characterize the fidelity of a specific gate. Here, we use the gate of interest (\text{I}, \text{X90}, \text{Y90}, \text{X180}) to construct the pseudo-identity gate (implemented as a \SI{40}{\nano\second} delay, four \text{X90} gates, four \text{Y90} gates, or two \text{X180} gates, respectively) which is interleaved between each Clifford gate in the random sequence.  
Experimental results are shown in \figref{fig:figure_fidelity}. For each circuit depth, measurements are performed for a large number (50) of different random sequences. It is clearly seen that the average gate fidelity exceeds \SI{99.96}{\percent}.

\subsubsection{5. QPT protocol and noise channels}

In our experiments, we performed std-QPT as shown in the End Matter with minor modifications allowing us to introduce different noise channels as desired. We detail the QPT protocol below.

1. \textbf{Active reset.} Each experiment begins with an active reset procedure to initialize the qubit in the ground state $|0\rangle$. Here, we start by measuring the qubit state with a readout pulse of \SI{2}{\micro\second}. After digitization, the state is encoded in the in-phase component according to an initial calibration similar to method described in \href{https://github.com/qua-platform/qua-libs/tree/main/Quantum-Control-Applications/Superconducting/Single-Fixed-Transmon/Use%20Case%202%20-%20Optimized%20readout%20with%20optimal%20weights}{https://github.com/qua-platform/qua-libs}, 
where we optimize the information obtained from the readout signal by deriving the optimal integration weights. The aim is to maximize the separation of the IQ blobs when the ground and excited state are measured. In a one-dimensional (1D) histogram, shown in the top right of Fig.5(b) in the main text
we can see that the peaks from the ground $|0\rangle$ (blue) and excited $|1\rangle$ (red) states are separated from each other along the in-phase quadrature (the histogram comprises one million single-shot measurements). This approach simplifies on-the-fly discrimination between the two quantum states by requiring only a single threshold $I_\text{Thresh}$ [vertical solid green line in Fig.5(b)].

Thus, based on the in-phase component, the qubit state is estimated. If the OPX+ detects the ground state $|0\rangle$, no pulse follows. If the excited state $|1\rangle$ is recognized, a \text{X180} pulse is executed to rotate it to the ground state $|0\rangle$, after which the measurement repeats.
As a result, the qubit is initialized in its ground state. It is worth noting that each measurement in std-QPT is followed by a buffer time $\SI{10}{\micro\second}$ before applying qubit pulses in order to avoid dephasing due to photon shot noise.

2. \textbf{Apply $U_1$.} Here $U_1$ is a set of known controlled unitary operations (\text{I}, $-\text{X90}$, $-\text{Y90}$, \text{X180}) applied to the ground $|0\rangle$ state. This prepares the qubit in different input states. As mentioned above, all single-qubit pulses have the same duration of \SI{40}{\nano\second}, with the real part of the pulse waveform shaped as a cosine envelope and the imaginary part corresponding to the DRAG correction, as shown in the left graph of Fig.5(c). 
For consistency, the \text{I} gate is implemented here as a \SI{40}{\nano\second} delay.

3. \textbf{Apply $\mathcal{E}$.} The quantum process under study (generally unknown). Experimental results for std-QPT, EM-QPT, and ML-QPT on the 24 single-qubit Clifford gates presented in the main text were obtained simultaneously by implementing a sweeping loop through the entire gate set listed in Table~\ref{tab:rotations}. The tomography data obtained for the identity operation \text{I} is then used in EM-QPT to define the error matrix $\tilde{\chi}^I$, as described in the main text.

4. \textbf{Apply $U_2$.} Here $U_2$ is a set of three measurement projectors (\text{I}, \text{X90}, \text{Y90}) to read out \textit{z}, \textit{y}, and \textit{x} components, respectively.
Unitary waveforms are implemented in the same way as for $U_1$.

5. \textbf{Readout operation.} For std-QPT the readout operation is a constant pulse with calibrated readout length and amplitude~$\text{A}_{r0}$ followed by $\SI{10}{\micro\second}$ buffer time. The measurement records consist of single-shot data (indicating whether the final state of the qubit is $|0\rangle$ or $|1\rangle$) obtained on the fly using the threshold $I_\text{Thresh}$, as well as the raw IQ data for further post-processing if necessary. 

The switching between state preparations, measurement projections, and processes under study, as well as the outer loop for averaging, is implemented via four nested QUA \texttt{for} loops directly on the OPX+. This enables precise real-time control, efficient parallelism, and strict preservation of experimental timing.
Considering the duration of the readout pulse, buffer time, and the duration of a single unitary rotation, we estimate the total time required for one QPT experiment across the whole Clifford group with 10,000 averages. Thus, the time for a single QPT is approximately $t_{QPT} \approx 10,000 \times 4\times 24\times 3 \times [\SI{2}{\micro\second} + \SI{10}{\micro\second} + \SI{0.12}{\micro\second} + \SI{2}{\micro\second} +\SI{10}{\micro\second}]=\SI{70}{\second}$, which enables us to acquire the 2D data plot shown in the main text in only 10 hours, using 15 QPT repetitions for reference error value. For comparison, the total duration of the RB experiment shown in \figref{fig:figure_fidelity} is approximately $\SI{435}{s}$, with data for all five plots acquired in parallel using only 100 averages per random seed.

\subsection{Appendix F: Model stability and training convergence}

\begin{figure}
\centering
\includegraphics[width=0.5\columnwidth]{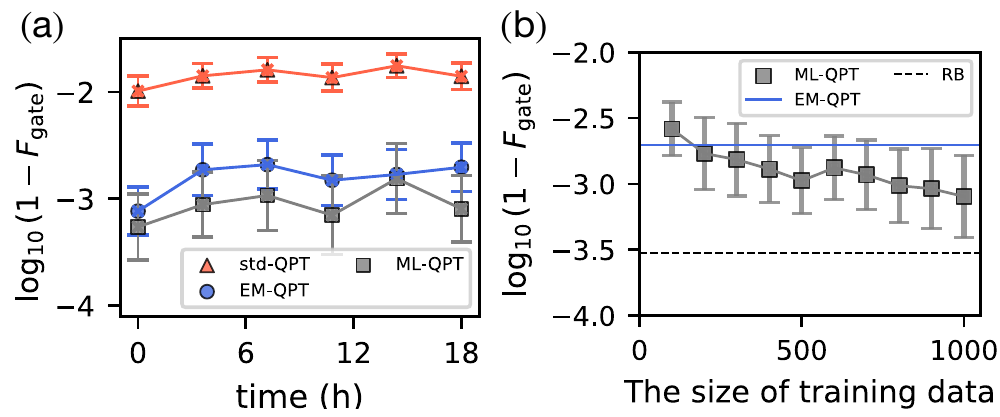}
\caption{\textbf{Stability and convergence.} 
(a) The QPT infidelity as a function of time (over 18 hours) across the three investigated methods. 
For ML-QPT, the digital twin is trained on a data set of $10^3$ error matrices. 
(b) The infidelity of ML-QPT as a function of the size of the training data, where the training data is randomly sampled from the full data set of $10^3$ error matrices. Each data point is averaged over the same measurement data set comprising $10^2$ \texttt{X} gates. For comparison, we include the average gate fidelity of EM-QPT (solid blue line) and the RB fidelity (dashed line).
Error bars in (a) and (b) correspond to the standard deviation of the QPT process for $10^2$ experimentally implemented \texttt{X} gates.
}
\label{fig:figure_app1}
\end{figure}

In this section, we analyze the stability of the trained model. A digital twin was first trained using a data set of $10^3$ samples. We then performed QPT six times over 18 hours, with each experiment comprising $10^2$ $\texttt{X}$-gate measurements and corresponding identity-gate measurements.
In \figpanel{fig:figure_app1}{a}, we compare the performance of the three different methods over time. The results demonstrate that ML-QPT exhibits stability comparable to EM-QPT while achieving even higher accuracy.

Figure~\figpanelNoPrefix{fig:figure_app1}{b} examines the training data requirements. Non-repeating subsets of $10^2$ to $10^3$ points were randomly sampled from the full data set ($N=10^3$) to train the digital twin. Each model was evaluated on the same $10^2$ $\texttt{X}$ gates via EM-QPT, with fidelities averaged over $10^2$ digital twin simulations of error matrices.
The infidelity converges to $\sim 10^{-3}$ when the training set exceeds $5 \times 10^2$ samples (error bars: standard deviation across $10^2$ $\texttt{X}$-gate measurements).

\subsection{Appendix G: High-precision and robustness of ML-QPT}

\begin{figure}
\centering
\includegraphics[width=0.55\columnwidth]{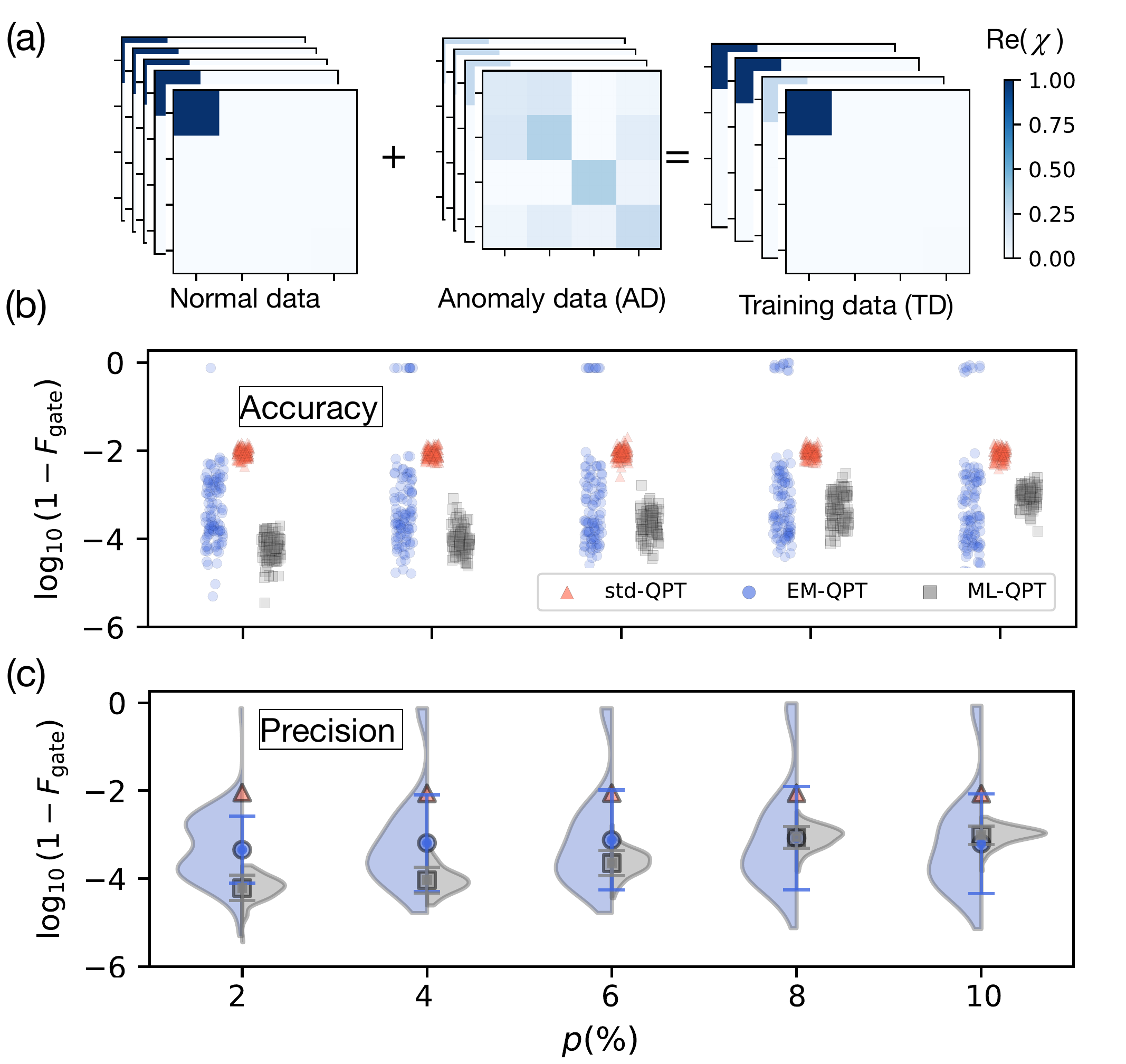}
\caption{\textbf{Performance benchmark with anomalies.}
(a) Schematic illustration of the training data, which consist of normal data and anomaly data represented by the real part of the $\chi$ matrix. The anomaly proportion $p$ is defined as the ratio of anomaly samples to the total training data.  
(b) Estimated gate infidelity as a function of $p$ for three different methods, with numerical experiments of $10^2$ single-qubit gates performed for each $p$.
(c) Example of gate fidelity estimation: the ML-QPT result is obtained by averaging over $10^2$ digital twins (grey distribution), compared with EM-QPT (blue distribution) and std-QPT (orange triangles). This demonstrates the higher-order moments captured by ML-QPT. Data points indicate mean values, and error bars represent one standard deviation. 
Note the ML-QPT is trained on datasets containing the same anomaly proportion $p$ in (b-c).  
Other parameters: total SPAM error $\lambda_{\text{tot}} = 0.02$, training data size $10^3$.}
\label{fig:figureSM4}
\end{figure}

In this section, we benchmark various QPT methods in terms of their robustness in the presence of experimental anomalies. In our setup, an anomaly is interpreted as an unexpected experimental condition, such as sudden heating, an instrument glitch, or some human error, resulting in completely uninterpretable measurements. We consider anomaly events to occur only in the identity process, since ML-QPT requires a large number of identity QPTs. This is also reasonable, as anomalies are more likely to arise in long experimental runs.

First, we numerically produce $10^3$ QPT measurements of the identity process for single-qubit gates in the presence of both coherent and incoherent errors with $\lambda_{\text{tot}} = 0.02$, based on the Qiskit simulator (similar to Fig.~2 of the main text). Then, we replace a proportion $p$ of the total training data with anomaly events in random order. Here, we define an anomaly event as the identity QPT experiments failing to produce reasonable readouts, leading to a completely different $\chi$ matrix from the normal identity process [see \figpanel{fig:figureSM4}{a} for an example]. In our simulation, for the sake of simplicity, we randomly sample the $\chi$ matrix as an anomaly event. Note that anomalies may also correspond to other abnormal behaviors in experiments. We encourage the reader to customize such scenarios using our open-source code (\href{https://github.com/huangtangy/EM-QPT}{https://github.com/huangtangy/EM-QPT}).

In \figpanel{fig:figureSM4}{b}, we use the same training data set with different amounts of anomalies (anomaly proportion $p$) to train the digital twin, and then compare the gate fidelities obtained from the trained generative model (ML-QPT), the real-time error-matrix approach (EM-QPT), and standard QPT (std-QPT) for $10^2$ randomly generated single-qubit gates, each with the same probability ($p$) of being anomalous. We see that EM-QPT fails to reconstruct the quantum process in some cases (the uppermost circular data points; the ones where an anomaly occurred), while ML-QPT demonstrates high precision and robustness, never failing like EM-QPT does. Note that std-QPT is independent of anomaly events since it is not relying on identity QPT.

Figure~\figpanelNoPrefix{fig:figureSM4}{c} shows the infidelity distribution as a function of anomaly proportion for estimating a single-qubit gate with the three QPT methods. Each distribution contains $10^2$ fidelites using error matrices based on real-time identity QPT (blue for EM-QPT) and digital twins (grey for ML-QPT). The precision of ML-QPT is significantly more robust than EM-QPT even with low-quality training data (when \SI{10}{\percent} of the data set consists of anomalies), although its accuracy slightly degrades once $p$ exceeds \SI{6}{\percent}. EM-QPT fails when the identity QPT itself is anomalous, leading to invalid gate-fidelity estimations [see the uppermost blue distribution in \figpanel{fig:figureSM4}{c}]. This result highlights the significant advantage of ML-QPT in practical experiments, particularly in the presence of anomaly events. Note that std-QPT only provides a single fidelity value without capturing statistical behavior (orange triangular data points).

ML-QPT shows considerable improvement in both precision and robustness compared to the other two methods. This improvement can be understood as follows:  
Firstly, the digital twin is capable of learning the essential features of error matrices for a given hardware, independent of the specific gate operation and resilient to a small portion of anomaly events, thereby enabling accurate and robust fidelity estimation compared to EM-QPT. In particular, EM-QPT fails when using an anomalous error matrix, whereas ML-QPT remains accurate even when up to \SI{10}{\percent} of the training set contains anomalous events [see \figpanel{fig:figureSM4}{b}].  
Secondly, ML-QPT provides high-precision gate fidelity estimation up to the second moment. Since the trained generative model can produce arbitrarily many digital twins of the error matrix, it enables reliable second-moment fidelity estimation for each gate, outperforming both EM-QPT and std-QPT [see \figpanel{fig:figureSM4}{c}].

Moreover, we here outline two possible directions for further improvement of ML-QPT.  
(i) One approach is to introduce a weighting factor (similar to the better-QPT method~\cite{blumekohout2024easybetterquantumprocess}) to bias the SPAM errors by incorporating prior knowledge about their relative contributions.  
(ii) A more technical route is to search for an optimal digital twin of the error matrices using advanced machine learning models, such as conditional generative adversarial networks (cGANs)~\cite{cGAN}. The core idea of this approach is to label high-quality error matrices within the training dataset, thereby guiding the model to generate superior digital twins. However, this cGAN-based strategy typically requires a significantly larger amount of training data and suffers from convergence difficulties, which makes it less practical for experimental implementation.

\subsection{Appendix H: Effects of biased SPAM errors}
\begin{figure}[h]
\centering
\includegraphics[width=0.5\columnwidth]{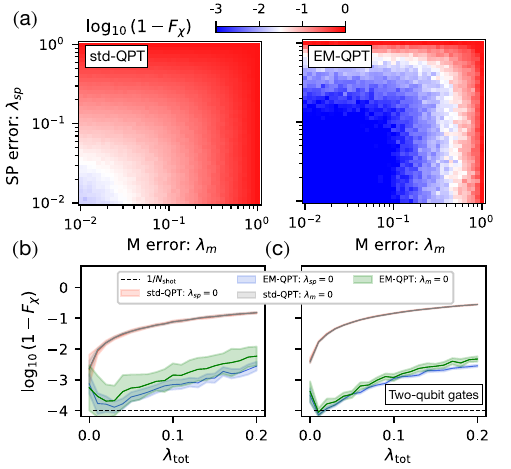}
\caption{ \textbf{Numerical results on biased SPAM errors.}
(a) Averaged process infidelity of single-qubit gates as a function of state-preparation (SP) and measurement (M) errors, obtained using std-QPT (left panel) and EM-QPT (right panel).  
(b–c) The logarithm of the process infidelity as a function of the total SPAM error $\lambda_{\rm tot} = \lambda_{\rm sp} + \lambda_{\rm m}$ for single- and two-qubit gates, respectively.  
Solid curves represent the average results over $10^2$ gate samples, while the shaded regions and error bars indicate one standard deviation.  
The horizontal dashed line denotes the statistical error $1/N_{\text{shot}}$, with $N_{\text{shot}} = 10^4$ shots.  
Note that both SPAM errors include coherent and incoherent components with identical error strengths (see the definition of error strength for coherent/incoherent error in the main text).
}
\label{fig:figure_reply}
\end{figure}

In our method, we construct noisy initial states and measurements by assuming that either the measurement or the state preparation is perfect:
\beq
\mathcal{J}[\{M_{\mu}\}, \tilde{\chi}^I] \rightarrow \{\bar{\rho}_{i}\}, \quad 
\mathcal{J}[ \{ {\rho}_{i}\}, \tilde{\chi}^I] \rightarrow \{\bar{M}_{\mu}\},
\eeq
using quantum state tomography and detector tomography, respectively.
This assumption in our method might lead to an overestimation of the process fidelity if the SPAM errors are extremely imbalanced, e.g., when the SP (or M) error is zero, i.e., $\mathcal{E}_{\rm sp} = \mathcal{I}$ (or $\mathcal{E}_{\rm m} = \mathcal{I}$), which corresponds to the case $p = 0$ or $p = 1$ in the ``better-QPT'' method proposed in Ref.~\cite{blumekohout2024easybetterquantumprocess}.  
To investigate this potential issue, we performed numerical simulations for EM-QPT with imbalanced SPAM errors. Note that we use SPAM error setups similar to those in Fig.~2 of the main text, with coherent and incoherent errors evenly assigned to either state preparation or measurement.
In \figpanel{fig:figure_reply}{a}, we present the infidelity of EM-QPT and std-QPT as functions of the SPAM errors.  
We observe that our method still outperforms std-QPT with consistently high fidelity, even under extreme conditions where $\lambda_{\rm sp} = 0$ or $\lambda_{\rm m} = 0$ (see panels (c-d)).
We interpret this result as follows: a high process fidelity can still be achieved because one can obtain the exact noisy states
$|\bar{\rho}_i\rrangle = |\tilde{\rho}_i\rrangle \to \tilde{\mathcal{E}}_{I}| {\rho}_i\rrangle$ 
when $\lambda_{\rm m} = 0$, even if $\llangle \bar{M}_i| \neq \llangle \tilde{M}_i|$, and vice versa. These numerical results show that EM-QPT enables high-precision fidelity estimation even under severe biased SPAM errors, demonstrating that noisy probes can be accurately reconstructed via the error matrix without loss of generality.

\end{document}